\documentclass[%
 reprint,
 amsmath,amssymb,
 aps,
prb,
superscriptaddress
]{revtex4-1}

\usepackage{graphicx}
\usepackage{dcolumn}
\usepackage{bm}

\begin{document}


\title{Diamond nanopillar arrays for quantum microscopy of neuronal signals}

\author{Liam Hanlon}
\email{Liam.Hanlon1@anu.edu.au}
\affiliation{Laser Physics Centre, Research School of Physics and Engineering, Australian National University, Canberra, ACT, 2601, Australia}
\author{Vini Gautam}
\affiliation{Eccles Institute of Neuroscience, John Curtin School of Medical Research, Australian National University, Canberra, ACT, 2601, Australia}
\author{James D. A. Wood}
\affiliation{Department of Physics, University of Basel, Klingelbergstrasse 82, Basel CH-4056, Switzerland.}
\author{Prithvi Reddy}
\affiliation{Laser Physics Centre, Research School of Physics and Engineering, Australian National University, Canberra, ACT, 2601, Australia}
\author{Michael S.J. Barson}
\affiliation{Laser Physics Centre, Research School of Physics and Engineering, Australian National University, Canberra, ACT, 2601, Australia}
\author{Marika Niihori}
\affiliation{Laser Physics Centre, Research School of Physics and Engineering, Australian National University, Canberra, ACT, 2601, Australia}
\author{Alexander R.J. Silalahi}
\affiliation{Research School of Biology, Australian National University, Canberra, ACT, 2601, Australia}
\author{Ben Corry}
\affiliation{Research School of Biology, Australian National University, Canberra, ACT, 2601, Australia}
\author{J{\"o}rg Wrachtrup}
\affiliation{$3^{rd}$ Institute of Physics and Research Centre SCOPE, University Stuttgart, Pfaffenwaldring 57, D-70550 Stuttgart, Germany}
\author{Matthew J. Sellars}
\affiliation{Laser Physics Centre, Research School of Physics and Engineering, Australian National University, Canberra, ACT, 2601, Australia}
\author{Vincent R. Daria}
\affiliation{Eccles Institute of Neuroscience, John Curtin School of Medical Research, Australian National University, Canberra, ACT, 2601, Australia}
\author{Patrick Maletinsky}
\affiliation{Department of Physics, University of Basel, Klingelbergstrasse 82, Basel CH-4056, Switzerland.}
\author{Gregory J. Stuart}
\affiliation{Eccles Institute of Neuroscience, John Curtin School of Medical Research, Australian National University, Canberra, ACT, 2601, Australia}
\author{Marcus W. Doherty}
\email{marcus.doherty@anu.edu.au}
\affiliation{Laser Physics Centre, Research School of Physics and Engineering, Australian National University, Canberra, ACT, 2601, Australia}

\keywords{nanopillars, Nitrogen-vacancy, neuro-imaging, neuromodelling, neurons.}

\date{\today}

\begin{abstract}
Modern neuroscience is currently limited in its capacity to perform long term, wide-field measurements of neuron electromagnetics with nanoscale resolution. Quantum microscopy using the nitrogen vacancy centre (NV) can provide a potential solution to this problem with electric and magnetic field sensing at nano-scale resolution and good biocompatibility. However, the performance of existing NV sensing technology does not allow for studies of small mammalian neurons yet. In this paper, we propose a solution to this problem by engineering NV quantum sensors in diamond nanopillar arrays. The pillars improve light collection efficiency by guiding excitation/emission light, which improves sensitivity. More importantly, they also improve the size of the signal at the NV by removing screening charges as well as coordinating the neuron growth to the tips of the pillars where the NV is located. Here, we provide a growth study to demonstrate coordinated neuron growth as well as the first simulation of nano-scopic neuron electric and magnetic fields to assess the enhancement provided by the nanopillar geometry. \textbf{Keywords:} nanopillars, Nitrogen-vacancy, neuro-imaging, neuromodelling, neurons.
\end{abstract}

\maketitle

Modern neuroscience is rapidly probing new frontiers in neuron electrophysiology. It is becoming clearer that in order to understand neuron excitability properly there are a variety of requirements that must be met. These requirements can be presented in four major areas. The first is millisecond or sub-millisecond temporal resolution, this allows for the measurement of action potential (AP) changes over fast timescales \cite{Peterka2011}. The second is nano-scale resolution, which allows probing of individual neuron compartments such as dendritic spines \cite{Novak2013a}. This challenge also requires high resolution at a single point as well as across a wide-field image of a whole neuron or several neurons in a network \cite{Peterka2011}. Thirdly the sensitivity of the probe must be exceptional, a sensor must be able to resolve millivolt changes in potential within sub millisecond timescales in order to sense the smallest signals produced by neurons \cite{Perron2009}. Finally the last major requirement is the stability of the sensor, the sensor must be easily applied to a neuron without altering its natural properties, where neurons can be measured many times without the sensor breaking or the cell dying. This allows for imaging of neuron changes over a long period of time, such as neuroplastic effects \cite{Massobrio2015a}.

There are a growing number of different techniques which can meet one or more of these requirements. Patch clamp technology allows nano-scale cross sectional spatial resolution and sub-millisecond temporal resolution at a single point \cite{Savtchenko2017a}. Coupled with scanning ion conductance microscopy, it can perform wide field imaging of neurons in a resting state \cite{Novak2013a}. Voltage sensitive fluorophores come in a variety of forms, some of which have been shown to be able to image nano-scale structures such as dendritic spines with high sensitivity \cite{Popovic2015}. However no single device has the capacity to meet all the listed requirements. The patch clamp technique can only measure APs at a single point on a neuron, removing the capacity of imaging propagation effects \cite{Peterka2011}. Voltage sensitive dyes can be difficult to use, requiring careful tailoring of the correct dye to the correct cell \cite{Peterka2011}. In addition to this, any type of fluorophore that could be injected into a cell has an inherent time limit before photobleaching renders the sensor inoperable or phototoxicity kills the cell being imaged \cite{Peterka2011}. 

Another often overlooked detail to consider when probing neurons at nano-scale resolution, this is how to best theoretically predict nano-scale neuron electromagnetics. Most spatial predictions of electromagnetics rely on variations of core conductor (CC) theory \cite{Savtchenko2017a, Woosley1985a}. The central assumption of this theory is that the density of ions inside and outside the neuron remain stationary during an AP. At micron distances from the membrane where the ions can form a stable equilibrium this assumption is true, making CC theory viable. However, at nano-scale distances from the membrane, ions flowing in and out of the neuron is precisely what generates the AP, thus the assumption that the ion density is stationary is not valid. 

In this paper we present a potential solution to nano-scale neurosensing with the application of the nitrogen vacancy centre (NV) \cite{Barry2016b, Hall2013}. The NV is a defect in a diamond lattice consisting of a substitutional nitrogen atom with a nearest neighbor vacancy \cite{Rondin2014}. The NV has unique spin dependent photo-dynamics that allow its electronic spin to be optically initialized and read out. This allows for the NVs electron spin resonance (ESR) to be measured using optically detected magnetic resonance (ODMR). The ESR frequency is dependent on perturbations from external electric \cite{Dolde2011a} and magnetic fields \cite{Rondin2014}, causing a resonance shift, which is measured by ODMR. Combining this with its atom-like size the NV can perform nano scale measurements of electric and magnetic fields.

The NV has been shown to have some of the best sensitivities and spatial resolutions for a room temperature sensor. For DC sensing, single NVs have been able to measure 2.8$\times10^9$~mV/m\cite{Dolde2011a} electric fields and 1.26~$\mu$T\cite{Rondin2014} magnetic fields in a 1~ms acquisition time; this is within a suitable range for neurosensing. In addition to its sensitivity, the NV has been shown to have sub-millisecond temporal resolution as well as spatial resolutions well into the nanometre scale \cite{Taylor2008a}. The NV itself is also a very stable atomic system, which does not suffer from photobleaching, allowing for long term imaging of a single sample \cite{Barry2016b}. In addition to its physical capabilities, the NV is situated in a diamond structure. Diamond has been shown to be biologically compatible, having been successfully utilized in the past as a substrate to grow biological samples \cite{May2012}. In fact, Barry et al. have been successful in measuring neuronal signals of worm axons by placing the axon on a flat diamond substrate with embedded NVs \cite{Barry2016b}. In addition to this, work by Karaveli et al. has demonstrated NV sensing of 20~mV changes in potential by utilizing it as a charge state sensor\cite{Karaveli2016}. However, as we move away from larger worm neurons towards the sensing of smaller mammalian neurons, the signal will also decrease beyond the sensitivity of the NV. Indeed, in this paper, our simulation results show that measuring signals from neurons 500~nm in radius \cite{Liewald2014a} is not possible using NV centres in unstructured diamond. 

\begin{figure*}[!ht]
            \centering
            \includegraphics[width = 0.9\textwidth]{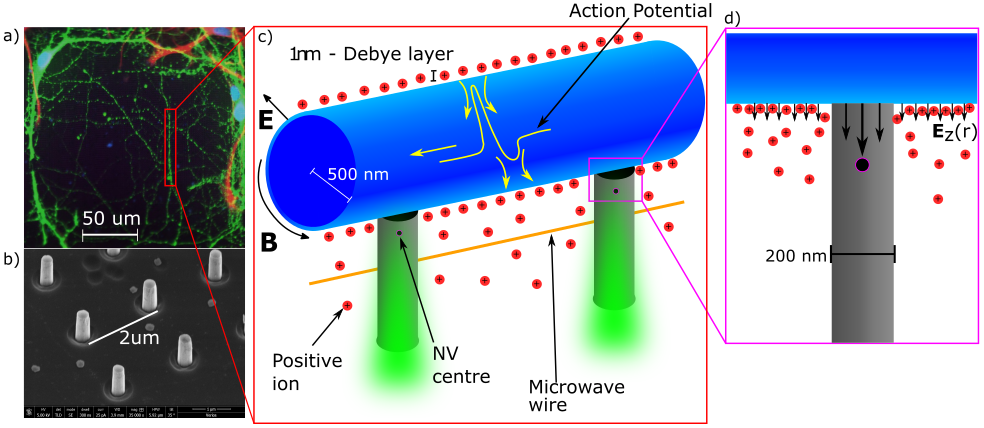}
            \caption{a) Confocal image of the stained neurons (green) growing on a bed of diamond nanopillars, which is visualized in the SEM image below b). c) Cutout of a neurite growing on the pillars. The panel shows the cylindrical neurite with positive ions forming the 1~nm thick Debye layer outside of it (negative ions not shown. Yellow arrows inside the cylinder indicate the current flow during an AP, which is depicted by the yellow line. The ion concentrations and current flows generate the electric (\textbf{E}) and magnetic fields (\textbf{B}), which are measured by the NVs situated in the grey diamond pillars. The sensing protocols use a green laser and microwaves to optically address the NV spin resonance. The pillars confine the laser light and direct the NV fluorescence. d) Illustration of how the pillar removes the Debye screening charges, increasing the radial electric field to a distance where the NV can sense the signal.}
            \label{NVstuff}
        \end{figure*}

Our solution to this significant problem was inspired by work done with diamond nano-optics \cite{Momenzadeh2015a, Babinec2010a} as well as neuron growth studies on Indium phosphate pillars by Gautam et al. \cite{Gautam2017a}. We seek to sense neuron signals with the fabrication of nanoscopic diamond pillars, each with an NV sensor embedded within it. The pillar geometry yields three advantages. Firstly, the shape of the pillars and the diamond's high refractive index guides the excitation and emission light in and out of the diamond without significantly illuminating the neurons themselves\cite{Momenzadeh2015a} (see figure \ref{NVstuff}c). This helps to improve the sensitivity of the NV by increasing the light collection efficiency by up to 5 times \cite{Momenzadeh2015a} and reduces the phototoxic effect on the neurons from extended illumination. The second advantage comes from the growth of the neurons on the pillars themselves. It has been demonstrated that pillar geometries act like a scaffold for neuron growth, guiding neurites in a single direction along the tips of the pillars \cite{Gautam2017a, Specht2004a} (figure \ref{NVstuff}a/c). This enhances the signal at the NV by coordinating neuron growth near the NVs themselves. The third advantage lies in removing the Debye layer of the neuron; the key element that is absent in CC theories are the effects of the Debye screening layer \cite{Savtchenko2017a}. The Debye layer is a build-up of ions on either side of the membrane due to the electrochemical forces acting on individual ions \cite{Hille2001}. The Debye length is typically around 1~nm \cite{Sykova2009a, Hille2001}, and is understood to greatly screen the electric potential external to the neuron \cite{Savtchenko2017a, Hille2001, Ridley2013}, reducing its magnitude to zero over the course of a few nm. A diamond pillar placed in contact with the neuron could remove the screening ions, increasing the propagation of the external electromagnetic field (figure \ref{NVstuff}d). It is our assumption that as long as the nanoscopic pillar is small enough compared to the microscopic cross-sectional area of the neuron, the removal of screening ions is unlikely to significantly alter the neuron's natural function. 

In this paper we provide a proof-of-concept study on the growth of neurons on a substrate of diamond nanopillars. The pillars are designed with a number of different diameters and pitches (distance from centre to centre of adjacent pillars) in order to assess which geometry is ideally suited for ordered neuron growth whilst maximizing photon collection efficiency for the sensing NVs. In order to predict how accurately NVs are able to detect changes in neuron electromagnetics we also conducted a simulation of an AP in a cylindrical axon with nanometre resolution. To the best of our knowledge, this will be the first neuron growth study using diamond nanopillars and the first simulation of neuron electric and magnetic fields solved for an entire axon with nano-scale resolution. Other simulations of this kind exist for simulating axon electric potentials \cite{Pods2013a, Lopreore2008a}, however this simulation will focus on electric and magnetic fields for NV sensing purposes. 

Following the work of Gautam et al. and other neuron growth studies \cite{Specht2004a, Gautam2017a, Franze2013a} on indium phosphide (InP) substrates, we fabricated an array of cylindrical diamond nanopillars following a recipe reported elsewhere \cite{Appel2016a}. We then cultured neurons on top of them, staining them and analyzing their growth using confocal microscopy. The method for the growth study is detailed in the supporting information. Once the neuron fluorescence was isolated on a diamond pillar patch, region of interest (ROI) processing was performed to obtain a metric for total neuron growth on a patch as a fraction of area. In addition to this we modified an image processing algorithm to determine the fraction of neurites coordinated with the rows or columns of the nanopillar grid. For each pillar patch we produce a mask, which identifies the ‘skeleton’ of each neurite. We then algorithmically traced the neurites in the mask and summed their alignment to a pillar line \cite{Lee1994a}. Thus a metric for coordinated or 'ordered' growth was obtained by analyzing the length of coordinated neurites against the total length of all neurites. The skeletonization process replaces neurites as well as cell bodies with lines to be measured. For this reason, it was unsuitable to use the algorithm as a metric for total growth as the method doesn't take into account the varying thickness of cell bodies or neurites. See supplementary material for more detail and image analysis software at \cite{Reddy2018}. Together, the total fluorescence and the ordering metric allow for an understanding of which diamond nanopillar geometries produce the ideal growth rates for sensing studies. 

The nanopillar arrays were arranged on patches labelled from 0 - 15 where each patch was 200$\times$200~$\mu$m. The pitch of individual pillars within a nanopillar patch ranged from 1 to 4~$\mu$m, with individual pillars having diameters of 200~nm and 350~nm and a height of 1~$\mu$m. We performed two repeats for every unique nanopillar patch geometry, however patches 0, 9 and 12 were removed from the results due to air bubbles, which covered the patch preventing growth. Growth statistics were obtained for total growth and ordered growth as functions of pitch, diameter and fractional separation factor ($p-d/p$). This was achieved by combining the results obtained from geometries with the same pitch, diameter or the fractional separation, respectively. Plots of total and ordered growth versus pitch are shown in figure \ref{fig:growthvspitch} and the remaining data can be found in the supplementary information. 

A key observation is that all patches showed non-negligible total growth. However, due to our small sample size, the standard error is such that no single nanopillar geometry exhibited statistically significant advantage for total growth. However, particular geometries had significantly larger growth ordering when considering pitch or fractional separation. These geometries had 2~$\mu$m pitch and either 200~nm or 350~nm diameters. When sampling for pitch, these geometries achieved 38\% ordering on average with a standard error of $\pm0.8$. There was no significant dependence on pillar diameter. This is likely due to the small range of diameters that were sampled, which was chosen to approximately match the range of diameters where the nanopillars maximize optical collection efficiency \cite{Momenzadeh2015a, Babinec2010a}, and thus sensitivity of the embedded NVs.  

\begin{figure}[!ht!]
            \centering
            \includegraphics[width = 0.43\textwidth]{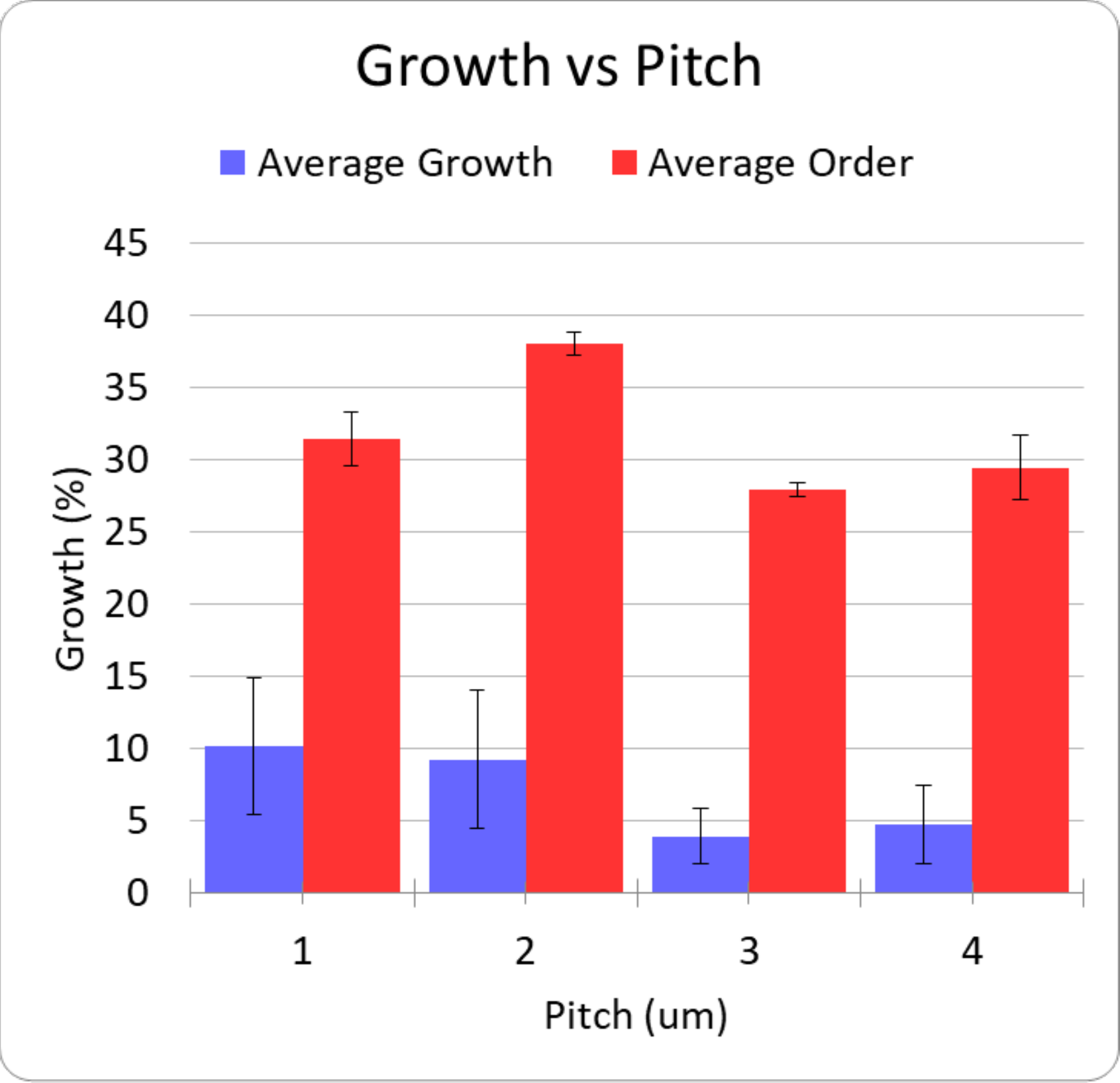}
            \caption{Table summary of growth as a function of pitch, averaged across all patches of the same pitch. Error bars indicate one standard error of the sample mean. There is a general trend towards higher ordered growth for 2~$\mu$m pitch pillars.}
            \label{fig:growthvspitch}
        \end{figure}

Our results are consistent with the previous results of Gautam et al. \cite{Gautam2017a}, who also observed that fractional separation is important to growth ordering. Our observation that growth ordering also significantly depends on pitch can be explained by our small range of pillar diameters, which meant that pitch was the dominant parameter in determining fractional separation. Having concluded this, future work is required to determine statistically significant total growth factors as well as finer pitch values to improve ordered growth. In addition to this Ca$^{2+}$ studies can help to determine neuronal activity on the diamond substrate. 

To model neuron electrophysiology we consider a neuron axon as an infinitely long cylinder. We then apply the coupled Poisson-Nernst-Planck (PNP) equations \cite{Lopreore2008a, Pods2013a, Corry2000} to this geometry: 

\begin{equation}\label{eqnPoissonMain}
    \epsilon \vec{\nabla^2} V(r) = -\rho(r) = -e \sum_{i = 1}^{M} z_i c_i(r) 
    \end{equation}

\begin{equation}\label{eqnDriftDiffuseMain}
    \frac{\partial{c_i}}{\partial{t}} = -\vec{\nabla} \cdot [D_i(\vec{\nabla} c_i + \mu_i c_i \vec{\nabla} V(r)]  
    \end{equation}
        
The Poisson equation (\ref{eqnPoissonMain}) utilises the charged ion concentration to solve for the potential and the Nernst-Planck equation (\ref{eqnDriftDiffuseMain}) utilises the electric potential to model ion concentrations in terms of the electrostatic and chemical forces that act on them. In this model, $c_i(r)$ is the ionic concentration, which is proportional to its charge density $\rho(r)$, and $V(r)$ is the electric potential. The mobility of the ions is given by $\mu_i$ ($\mu_i = \frac{D_i^2k_bT}{ze}$), $k_b$ is the Boltzmann constant, $T$ the temperature, $e$ the electric charge and $z_i$ the ion valency. The index $i$ denotes which ionic species is being studied, so the total potential will be the solution to the coupled equation, summed over all the participating ion species (up to the total, $M$). Although this model can be expanded to include any number of different ion species, for ease of calculation, we only consider the monovalent species: sodium (Na$^+$), potassium (K$^+$), chlorine (Cl$^-$) and negatively charged proteins produced by the neuron (OA$^-$). Although other ions do exist, their concentrations are considered low enough to be neglected. Additionally, the positive ions and the negative ions are grouped into two sources by defining averaged diffusion constants. We use 2D axisymmetric arguments to simplify the calculation, as well as a travelling wave assumption to convert the time dependent PNP equations to axial space dependent equations. From the PNP solutions we obtain the electric potential and the ion concentrations. The potential can be converted to an electric field ($\vec{E} = -\nabla V$), and the ion concentrations can be converted to a current density in order to solve for the magnetic field via Ampere's law.

To solve the PNP equations, the boundary conditions must be clearly stated and studied for both the ion concentrations and the electric potential. Far from the neuron radially, the boundary conditions are straightforward as the electric potential must go to zero, and the ion concentrations must reflect this with a stable equilibrium. Axially, far from the AP along the neuron, we expect the ion concentrations and the electric potential to reach a constant equilibrium corresponding to the resting potential (-68~mV)-thus, the derivative of the potential and flux must be zero in this region. At the membrane, we can derive a Neumann boundary condition using Gauss' law. In this derivation, the charge density is expressed in terms of the radial current and the transmembrane potential, both of which are obtained from the standard Hodgkin-Huxley (HH) \cite{HH1952, Zandt2011a, Lopreore2008a} equations of neuron APs. To obtain the membrane magnetic field boundary condition we apply Ampere's law where  the axial current is derived from the HH equations. The full derivation of our model as well as parameter values can be found in the supplementary information.

\begin{figure}[!ht]
            \centering
            \includegraphics[width = 0.45\textwidth]{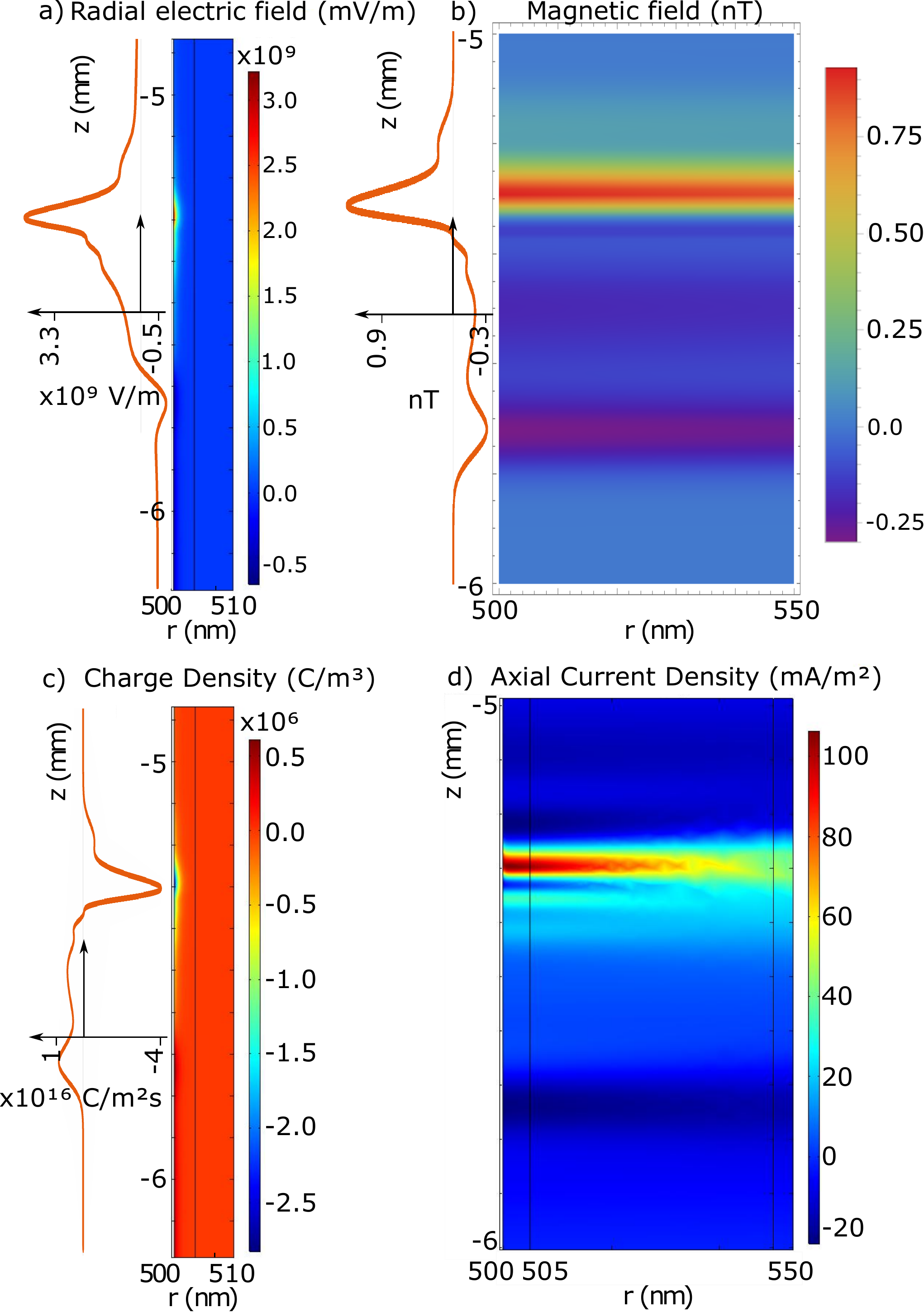}
            \caption{Simulation results for the a) electric field, b) magnetic field, c) charge density and d) axial current density. On the left of a), b) and c), the electric field, magnetic field and positive ion flux at the membrane are sketched (orange line), respectively. This demonstrates how these quantities longitudinally propagate with the neuron signal}
            \label{Result:2D}
        \end{figure}

\begin{figure}[!ht]
            \centering
            \includegraphics[width = 0.45\textwidth]{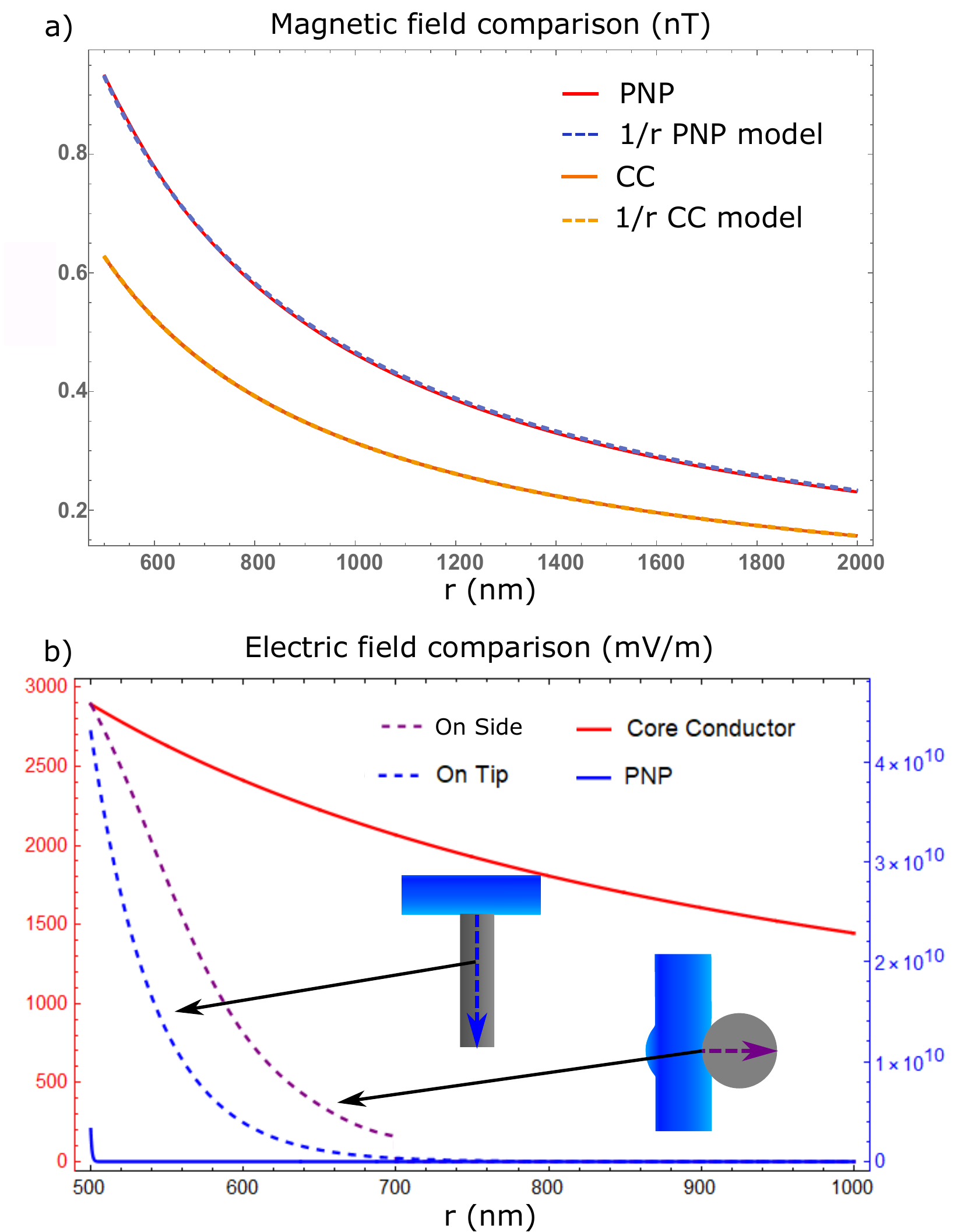}
            \caption{Radial 1D plots of the a) magnetic and b) electric fields taken from the peak of the AP wave. The magnetic field plots feature the PNP solution, the CC solution derived from Woosley \cite{Woosley1985a} as well as 1/r model fits for both. The electric field plots contain solutions to the PNP, CC models and the Laplace solutions for the electric field in the diamond when the neuron is in contact along the tip as well as along the side of the diamond pillar. The red axis is for the CC model and the blue axis is for the other two.}
            \label{Result:1D}
        \end{figure}

Surface plots of the results are displayed in figure \ref{Result:2D}, in the plots the 500~nm mark is the point defining the external membrane radius. The orange lines are the travelling wave signal moving axially along the neuron. These lines are taken from the membrane boundary condition solutions for their respective quantities and are plotted with more detail in the supplementary section. There are also radial line graphs of the solutions, which depict the electric and magnetic fields at the peak of the AP wave as well as 1/r model fits for the magnetic fields (figure \ref{Result:1D}). These plots also feature a CC solution calculated from the equations presented in Woosley et al. \cite{Woosley1985a}, but altered to match the parameters of the mammalian neuron considered in this study.  

As mentioned, the key result of the PNP calculation that is absent in CC theory is the effects of the Debye screening layer \cite{Savtchenko2017a}. The Debye screening is most evident in the electric field and charge density solutions. In figure \ref{Result:2D}a there is a radial electric field generated by the resting charge concentration at the membrane level far from the AP ($\approx -1.4\times10^8$~mV/m), along with a significant change at the peak of the AP wave to $\approx 3.3\times10^9$~mV/m. However, despite the dramatic change in electric field at the membrane level, the Debye screening reduces the magnitude of these fields to zero over approximately 3~nm (solid blue line in figure \ref{Result:1D}b). This results in a much larger field at the membrane level, but with a much more rapid decay of the radial electric field when compared to the CC result. This result is also similar in the charge density solutions (figure \ref{Result:2D}c) where almost everywhere outside the neuron the ion concentrations approach their bulk values, creating a zero charge density. At the AP region within the Debye layer the charge density decreases to as low as $\approx -3\times10^6$~C/m$^3$ due to the positive charge being transferred from outside to inside the neuron, increasing the potential and electric fields. Away from the AP wave, the charge density increases slightly to $\approx 0.5\times10^6$~C/m$^3$, re-forming the positive charge Debye layer in response to the internal negative charge present during the neuron resting conditions.

Debye screening plays less of an effect in the propagation of magnetic fields (figure \ref{Result:2D}b). Furthermore, the external current density result suggests that the external axial current runs parallel to the internal axial current, resulting in a magnetic field that is slightly enhanced outside the membrane (figure \ref{Result:2D}d). However, this current density is extremely small compared to the internal current ($\approx 100$~mA/m$^2$ at its peak) thus, the total magnetic field enhancement is negligible. The size and direction of the axial current density suggests that radial current densities are more responsible for the re-establishment of the neuron resting potential. The decay rate of the magnetic field has a 1/r dependence similar to the CC theory, which decays to zero over the range of $\approx 2$~$\mu$m. The major difference between the two theories is the larger membrane magnetic field in the PNP model. 

Critically, the magnetic field reaches a maximum of 0.95~nT at the membrane boundary. Whether the calculation is done using a full PNP simulation, CC theory\cite{Woosley1985a} or even by approximating the neuron as a current carrying wire \cite{Barry2016b}, the results tend to suggest that a mammalian neuron can only produce a magnetic field that is less than a nT in magnitude. This signal is too small to be detected by an NV within the millisecond timescale of the neuron signal\cite{Rondin2014}. Even with the improvement from the pillar geometry, a five-fold improvement to the light collection efficiency provided by the pillars \cite{Momenzadeh2015a} only creates an $\approx$~2 fold improvement to the magnetic field sensitivity, which still won't allow for mammalian neuron AP sensing \cite{Taylor2008a}. For electric field sensing, although the magnitude of the field is considerably larger than the minimum detectable electric field, the difficulty lies in placing an NV within the Debye layer external to the membrane. The closest range an NV can be placed to the surface of a diamond is $\approx 5$~nm whilst maintaining reasonable coherence and stability \cite{DeOliveira2017}. If the diamond tip is in contact with the neuron, at this distance the field will decay to zero.  

The analysis thus far suggests that NV sensing of neuron electric and magnetic fields is impossible. However one concept that has not been considered so far is the effect of a diamond nanopillar on the screening charge and current densities. As shown above, the effects of screening currents are small, and so the presence of the pillar will have a negligible effect on the magnetic field. However, the screening charge has profound effects on the electric field. By removing the screening charge between the neuron and the pillar tip (via good contact) and accounting for the much lower dielectric screening in the diamond ($\epsilon \approx6$) compared to the surrounding water ($\epsilon \approx 80$), we expect the electric field inside the pillar to be much larger. 

To model this enhancement, we solved Laplace's equation inside the pillar by assuming that the charge inside the neuron is unperturbed by the presence of the pillar and that Debye screening fixes the electric potential to be zero everywhere on the diamond surface where the neuron isn't in contact. Note that this ignores the small region at the edge of the contact area where, within the Debye layer extending from the neuron surface, the potential is non-zero at the diamond surface. We expect this to be a good approximation as long as the contact area dimensions are much larger than the Debye length, but not so large that the contact with the pillar changes the function of the neuron. This simulation was done for two different co-ordinations of the pillar and neuron with a pillar 200~nm diameter. One where the neuron is on top of the pillar and in contact with the pillar's complete circular top surface (area = $\pi(d/2)^2$), and the another where the neuron is on the side of the pillar near its tip, and has a square contact area of the same size. 

Plots of selected results are shown in figure \ref{Result:1D}b (see supplementary material for the full results and derivation). Specifically, these results are for the electric field magnitude in the pillar along a line extending from the centre of the contact area along/through the central axis of the pillar for the on-top (blue dashed line) and on-side (purple dashed line) co-ordination. The electric field of the on-top coordination has an analytic solution with an exponential decay determined by a decay constant of $\approx 4.8/d$. For a NV depth of $\approx 5$~nm and a pillar diameter much larger than this depth, this implies that an electric field as high as $\approx 3.8\times10^{10}$~mV/m will occur at the NV. For the on-side coordination, the electric field is predicted to be larger than for the on-top co-ordination. Indeed, the results show that at 100~nm away from the neuron membrane (i.e. the central axis of the pillar) electric fields as high as $\approx 1.02\times10^{10}$~mV/m will occur. The larger field that arises when the neuron is on-side is due to the curvature of the contact area. This curvature implies that the distance from a point in the pillar to a charge on the neuron surface is on-average smaller than for the flat contact area when the neuron is on top. This leads to a larger electric field for the on-side coordination. 

Given the NV electric field sensitivity mentioned above and the geometries stated, this modelling shows that the NV can easily detect APs within the signal timescale when the neuron is in both the on-top and on-side locations. However, this analysis also highlights the importance of having good contact between the neuron and the nanopillar in the region of the NV and that the location of this contact influences the electric field at the NV centre. Previous neuron growth studies show that neurons tend to grow towards the tip of a pillar and can form contact with either the top or side, with contact on the side being more common \cite{Hanson2012, Gautam2017a}. Future growth studies should seek to confirm this in a diamond substrate. In our results, it was not possible to determine the vertical position of the neuron in relation to the pillars nor the level of contact between the neuron and the pillar. Super-resolution confocal microscopy or SEM studies should be performed in future to examine the precise vertical position of the neurons.  
        
NV sensing of neuron signals has the capacity to provide a wealth of new information towards the understanding of neuron signalling for mammalian neurons. Our simulation results are essential for understanding what to expect from neuron signals and have the potential improve our understanding of how neurons function at the most fundamental level. The interesting result from our simulation lies in what external fields are measurable. Our calculations indicate that mammalian neuron magnetic fields are too small to be detected by NV centres within the timescale of the neuronal signal. However, electric field sensing is possible due to the large signal and the potential capacity of a nanopillar to remove the screening ions. The pillar geometry also improves NV optical collection efficiency and coordinates neuron growth to improve NV positioning with a neurite. Indeed, growth ordering up to 38\% was demonstrated, which implies that a high proportion of the neurite length is in a promising position for sensing.

This proof-of-concept simulation and growth study indicates the need to perform further simulations of NV pillars in a full interior and exterior solution of neuron electromagnetics to confirm the results presented in this paper. A larger scale growth study will also help confirm the ideal pillar dimensions for total growth and growth ordering. Additionally, further studies of the neurite's position relative to the pillar as well as the level of contact are also required. These studies will help determine the positioning of NVs within the pillars that on-average optimizes coordination with the neuron and thus neuron sensing.

\subsection*{Acknowledgements}

The authors acknowledge the support from the Australian Research Council (DP 170102735 and FT 130100781). We also acknowledge access to Australian National Fabrication Facility (ANFF ACT node) as well as Microscopy Australia at the Centre for Advanced Microscopy (CAM), both at the Australian National University.
A. Silalahi also acknowledges the support from the Australian Government Endeavour Fellowship. Finally the authors wish to thank Dr Andrea Zappe for her critical comments on the manuscript.

\clearpage

\section*{Supplementary Information}
\subsection{Growth study details}

Although the specific mechanism for the ordered growth is still unknown, a strong hypothesis is that ordered growth is centred around mechanosensitive structures in the neuron cytoskeleton \cite{Franze2013a}. During neuron growth on a protein layered substrate (e.g. laminin on a diamond pillar), neurons express proteins such as integrin molecules in all directions, which bind to the laminin in the extra-cellular matrix forming a new protein complex. This protein complex then binds to actin on the neuron cytoskeleton and mechanosensitive ion channels on the cytoskeleton initiate various mechanotransduction pathways, which encourages cellular growth at the point of the binding. As this process repeats itself, a regular line of pillar structures can encourage the process to continue in a line, thus producing directionally ordered growth. In this way, external forces on mechanically sensitive ion channels direct neuron growth in response to physical cues such as a diamond nanopillar \cite{Franze2013a}. One result of this is that there is a distance in which, ordered growth is maximized. The protein complexes produced in the growth process have a finite size, creating a range where the neuron is mechanosensitive. If the distance between the pillars is larger than this, then there won't be a connection of growth between adjacent pillars. However if the distance is too small, then the neurons could potentially grow in any direction towards an adjacent pillar (e.g. diagonally instead of vertically or horizontally) as long as it is within the range produced by the protein complex, thus producing no ordered growth at all similar to a flat substrate. This necessitates the need for a growth study to find the ideal pillar geometry that matches the growth mechanism.

\begin{figure*}[!ht!]
            \centering
            \includegraphics[width = 0.7\textwidth]{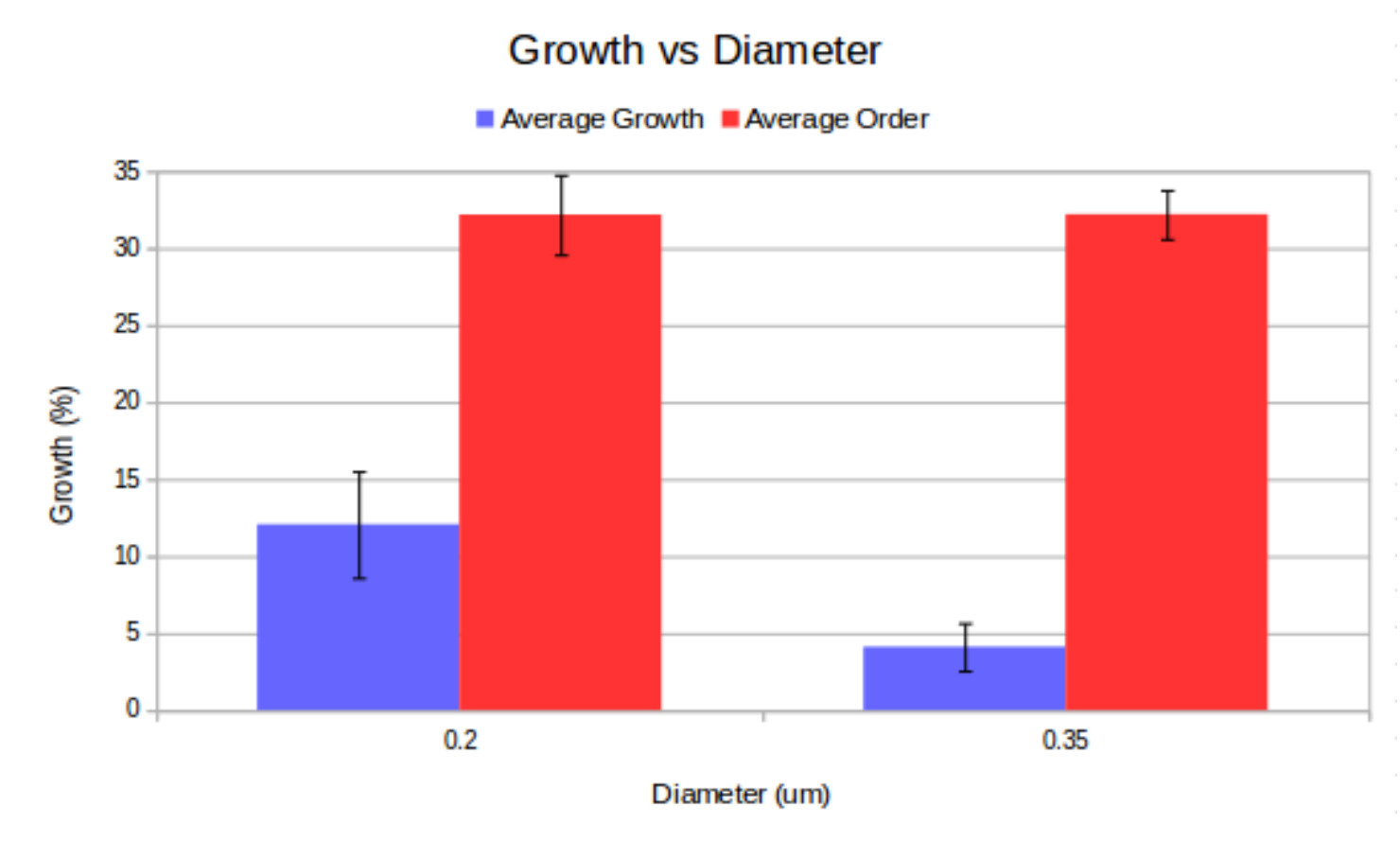}
            \caption{Table summary of growth as a function of diameter, averaged across all patches of the same diameters. Error bars indicate one standard error of the sample mean. There is no statistical dependence of diameter on ordered or total growth}
            \label{fig:growthvsdiameter}
        \end{figure*}

\begin{figure*}[!ht!]
            \centering
            \includegraphics[width = 0.7\textwidth]{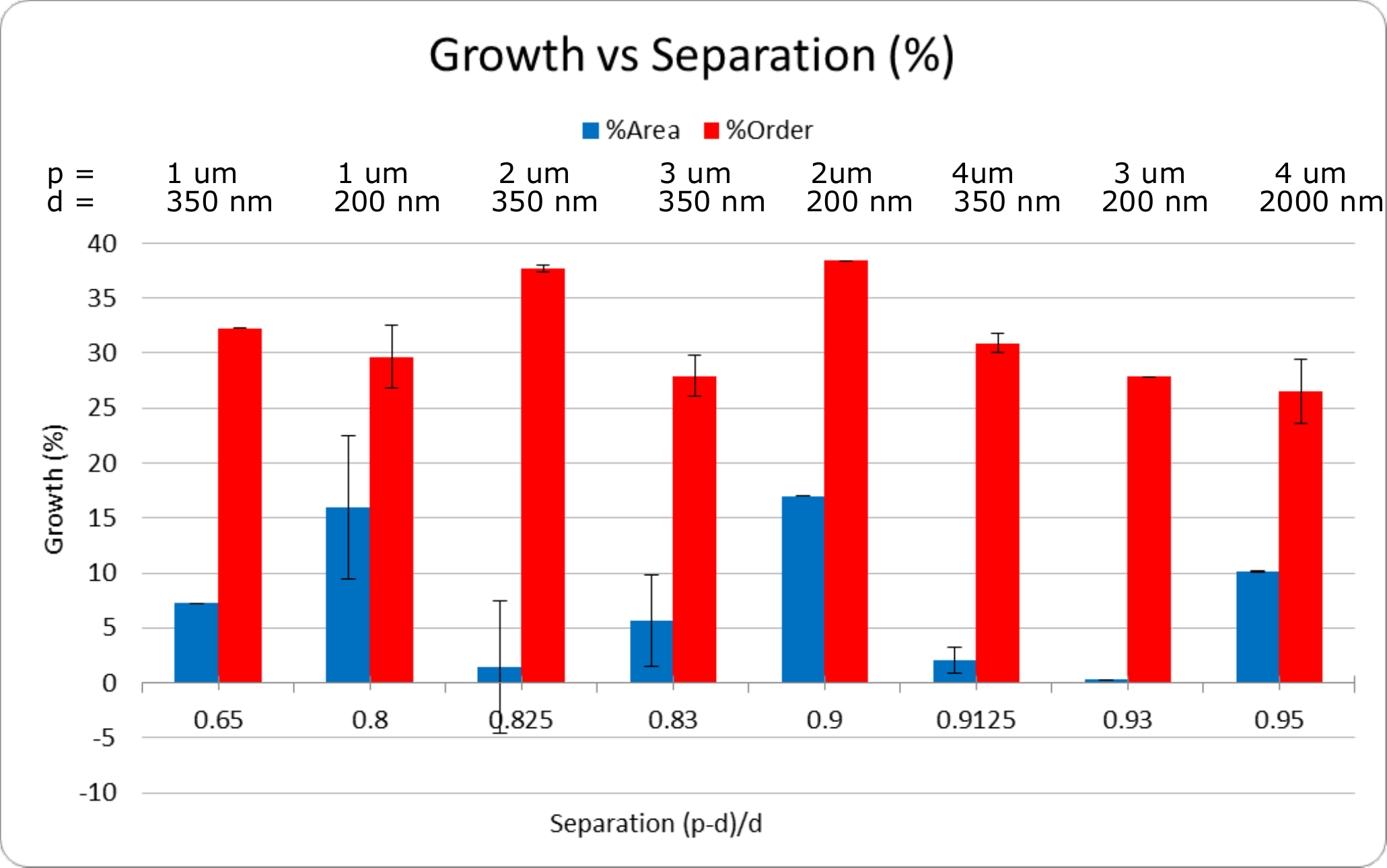}
            \caption{Table summary of growth as a function of the fractional separation, averaged across all patches of the same separation. The pitch and diameter for each separation is displayed above. Error bars indicate one standard error of the sample mean. The results show that high ordering occurs around 2~$\mu$m pitches, which is a similar result when averaged over pitch alone.}
            \label{fig:growthvsseparation}
        \end{figure*}

The dyes used for the growth study are listed in table \ref{table:dyes}. The table also includes the excitation and emission wavelengths for the various dyes as well as the Raman line we used to image the diamond itself. There was some overlap between the diamond fluorescence and the Glia dye, however as the diamond imaging was only used to find the pillar patch neurons were growing on, this did not affect the overall results. With the above exceptions, the cell labelling and confocal microscopy techniques were exactly the same as preformed in by Gautam et al. \cite{Gautam2017a}. Table \ref{table:GrowthResult} is the full details of the growth study result. Figures \ref{fig:growthvsdiameter} and \ref{fig:growthvsseparation} are a summary of the total growth and ordered growth averaged over the patches with the same diameter or fractional separation factor. 

The pillar pitches chosen following work from Gautam et al. \cite{Gautam2017a}, where distances between pillars were chosen to maximize the growth via the binding protein complexes. The diameters and heights were chosen by following work from Momenzadeh et al.\cite{Momenzadeh2015a}. In their work, they calculated the size and shape of pillars which maximizes light collection efficiency for shallow NV implantation. The general principle is that the pillar acts like a waveguide, whose size and shape matches to fundamental HE (hybrid electric) modes, guiding light out below the pillar into the detection system. Pillars 1~$\mu$m in height with 200 and 350~nm diameters are ideally shaped to maximize the number of fundamental modes guided in and out of the pillar.     
    
\begin{table}[!ht]
\begin{tabular}{|l|l|l|l|}
\hline
Technique & \begin{tabular}[c]{@{}l@{}} Type \\ imaged\end{tabular} & \begin{tabular}[c]{@{}l@{}}Laser \\ excitation\\ (nm)\end{tabular} & \begin{tabular}[c]{@{}l@{}}Emission \\ band \\ (nm)\end{tabular} \\ \hline
Alexa Fluor 405 & Cell Nuclei                                                       & 405                                                             & 450/50                                                            \\ \hline
Alexa Fluor 488 & Neuron                                                            & 488                                                             & 525/50                                                            \\ \hline
Alexa Fluor 594 & Glia                                                & 561                                                             & 595/50                                                            \\ \hline
Raman           & Diamond                                                 & 561                                                             & 585/65                                                            \\ \hline
\end{tabular}
\caption{Table of fluorescent components used in the confocal microscopy and the structures being imaged with them}
\label{table:dyes}
\end{table}

\begin{table*}[!ht]
\begin{tabular}{|l|l|l|l|l|l|l|l|l|}
\hline
Patch & \begin{tabular}[c]{@{}l@{}}Pitch \\ (p)\end{tabular} & \begin{tabular}[c]{@{}l@{}}Diameter\\ (d)\end{tabular} & \begin{tabular}[c]{@{}l@{}}Distance\\ ratio\\ (p-d)/d\end{tabular} & \begin{tabular}[c]{@{}l@{}}Volume\\ ratio\\ $\pi r^2/p^2$\end{tabular} & \begin{tabular}[c]{@{}l@{}}Total\\ Growth\\ \%Area\end{tabular} & \begin{tabular}[c]{@{}l@{}}Total \\ neurite\\ $T_i$ ($\mu$m)\end{tabular} & \begin{tabular}[c]{@{}l@{}}Ordered \\ neurite\\ $T_o$ ($\mu$m)\end{tabular} & \begin{tabular}[c]{@{}l@{}}Ordered\\ growth ratio\\ $\frac{T_i}{T_o}(\%)$\end{tabular} \\ \hline
0     & 1                                                    & 0.2                                                    & 0.8                                                                & 0.031                                                                         & 0                                                             & 0                                                                 & 0                                                                  & 0                                                                              \\ \hline
1     & 1                                                    & 0.2                                                    & 0.8                                                                & 0.031                                                                         & 15.9                                                            & 3762.5                                                                 & 1116.4                                                                  & 29.7                                                                              \\ \hline
2     & 1                                                    & 0.35                                                   & 0.65                                                               & 0.096                                                                         & 13.7                                                            & 5075.4                                                                 & 1494.5                                                                  & 29.4                                                                              \\ \hline
3     & 1                                                    & 0.35                                                   & 0.65                                                               & 0.096                                                                         & 0.7                                                             & 331.7                                                                  & 116.3                                                                   & 35.1                                                                              \\ \hline
4     & 2                                                    & 0.2                                                    & 0.9                                                                & 0.008                                                                         & 12.8                                                            & 3014.8                                                                 & 1148.4                                                                  & 38.1                                                                              \\ \hline
5     & 2                                                    & 0.2                                                    & 0.9                                                                & 0.008                                                                         & 21.1                                                            & 4119                                                                   & 1593.3                                                                  & 38.7                                                                              \\ \hline
6     & 2                                                    & 0.35                                                   & 0.825                                                              & 0.024                                                                         & 2.2                                                             & 796.4                                                                  & 315.0                                                                   & 39.6                                                                              \\ \hline
7     & 2                                                    & 0.35                                                   & 0.825                                                              & 0.024                                                                         & 0.8                                                             & 475.1                                                                  & 169.9                                                                   & 35.8                                                                              \\ \hline
8     & 3                                                    & 0.2                                                    & 0.933                                                              & 0.003                                                                         & 0.3                                                             & 341.8                                                                  & 95.2                                                                    & 27.9                                                                              \\ \hline
9     & 3                                                    & 0.2                                                    & 0.933                                                              & 0.003                                                                         & 0                                                            & 0                                                                 & 0                                                                   & 0                                                                              \\ \hline
10    & 3                                                    & 0.35                                                   & 0.833                                                              & 0.011                                                                         & 6.9                                                             & 1316.5                                                                 & 356.0                                                                   & 27.0                                                                              \\ \hline
11    & 3                                                    & 0.35                                                   & 0.833                                                              & 0.011                                                                         & 4.5                                                             & 1503.7                                                                 & 433.2                                                                   & 28.8                                                                              \\ \hline
12    & 4                                                    & 0.2                                                    & 0.95                                                               & 0.002                                                                         & 0                                                             & 0                                                                    & 0                                                                     & 0                                                                               \\ \hline
13    & 4                                                    & 0.2                                                    & 0.95                                                               & 0.002                                                                         & 10.1                                                            & 1965.5                                                                 & 521.5                                                                   & 26.5                                                                              \\ \hline
14    & 4                                                    & 0.35                                                   & 0.913                                                              & 0.006                                                                         & 2.0                                                             & 695.9                                                                  & 235.5                                                                   & 33.8                                                                              \\ \hline
15    & 4                                                    & 0.35                                                   & 0.913                                                              & 0.006                                                                         & 2.1                                                             & 1052.5                                                                 & 294                                                                     & 27.9 \\ \hline                                                                             
\end{tabular} 
\caption{Table of the growth results, including patch label, total growth, ordered growth, fractional separation and volume ratios. Note that patches 0, 9 and 12 had problems with the growth, requiring their data results to be removed from the published results and analysis.}
\label{table:GrowthResult}
\end{table*}

\subsection{Ordered growth analysis}
The orientation and length of each neuron can be determined by calculating the centre line of each neurite. A binary mask showing neurite centre lines was calculated by applying a skeletonisation algorithm \cite{Lee1994a}. This algorithm reduces every neurite and cell body in the image to a single pixel line without changing the overall structure of the image. The binary mask was convolved by a 3x3 kernel, such that neurite ends, midpoints, and intersections can be uniquely identified. Treating the resulting image as an undirected graph, we can parameterize each neurite by searching the graph for connected lines of pixels. Our search algorithm starts from any endpoint or intersection and traverses connected pixels until it finds another endpoint. Each set of connected pixels is called a path. This process is repeated for each endpoint until all paths are identified. Each path parametrizes the centerline of a neurite. We perform the following line integral along these parameterized centre lines to estimate how well the neurite aligns to a pillar line:

\begin{equation}
    T_i = \sum_{i=1}^{L} \int_{0}^{T} |\frac{\partial}{\partial t} \vec{f}_{i}| dt 
\end{equation}

\begin{equation}
    T_o = \sum_{i=1}^{L} \int_{0}^{T} \Theta(t) |\frac{\partial}{\partial t} \vec{f}_{i}| dt 
\end{equation}

where $T_i$ is the total length of all the summed neurites (paths), $T_o$ is the total length of the aligned (or ordered) neurites and $\vec{f}_{i}$ is the neurite's vector component parametrized by the length $t$:

\begin{equation}
    \vec{f}_{i} = x(t)\hat{x}+ y(t)\hat{y} 
\end{equation}

where the coordinate vectors $\hat{x}$ and $\hat{y}$ are chosen to coincide with the directions of the rows and columns of the nanopillar array.

The $\Theta(t)$ term represents a piece-wise function, which defines alignment by measuring the angle between the neurite vector component and the vector components of the pillar lines:

\[\Theta(t) = \begin{cases} 
      1, & arccos\Bigg(\frac{\frac{\partial \vec{f}}{\partial t}\cdot \hat{u}}{|\frac{\partial \vec{f}}{\partial t}|}\Bigg) \leq \frac{\pi}{36}  \\
      0, & arccos\Bigg(\frac{\frac{\partial \vec{f}}{\partial t}\cdot \hat{u}}{|\frac{\partial \overline{f}}{\partial t}|}\Bigg) > \frac{\pi}{36}
   \end{cases}
\]

where $\hat{u}=\hat{x}$ or $\hat{y}$. The principle is that the angle between a neurite vector components and a vertical ($\hat{y}$) or horizontal ($\hat{x}$) line of pillars is measured, if that angle is larger than our defined value ($\frac{\pi}{36}$) then the neurite is considered unaligned with the pillars and discarded. This process is repeated and summed for all neurite vector components ($T_o)$ and divided by the total integrated length of all neurites ($T_i$) to obtain our order ratio. 

Before applying this procedure we preprocessed the raw confocal image of each patch to isolate the the neuron fluorescence. In particular, we started by masking large cell bodies, such as glial cells, either by hand or using a disk shaped structuring element. We then filtered in the color space to extract the fluorescence from neuron neurites and applied a intensity threshold to filter out residual fluorescence from sources other than neurons. The resolution and quality of confocal scan is the most important factor determining the error in our alignment estimates as well as the ROI processing for total growth. Some factors are mitigated by preprocessing and denoising our image. For example, background fluorescence from other features is mitigated by our preprocessing steps. However, since we needed to apply an intensity threshold over the image, we also ignored low intensity fluorescence from neurites. Another shortcoming of our approach is that it does not correct for discontinuities in the neurites. This is not a problem when the length of each segment of a neurite is longer than the gaps. If a neurite shows up as a line of disconnected dots it’s alignment will not be measured correctly as the dots have no directionality, adding to the total neurite length as noise. This error can be quantified by measuring the number of short and singleton paths relative to the total length of detected neuron.

\begin{figure*}[!ht!]
            \centering
            \includegraphics[width = 0.9\textwidth]{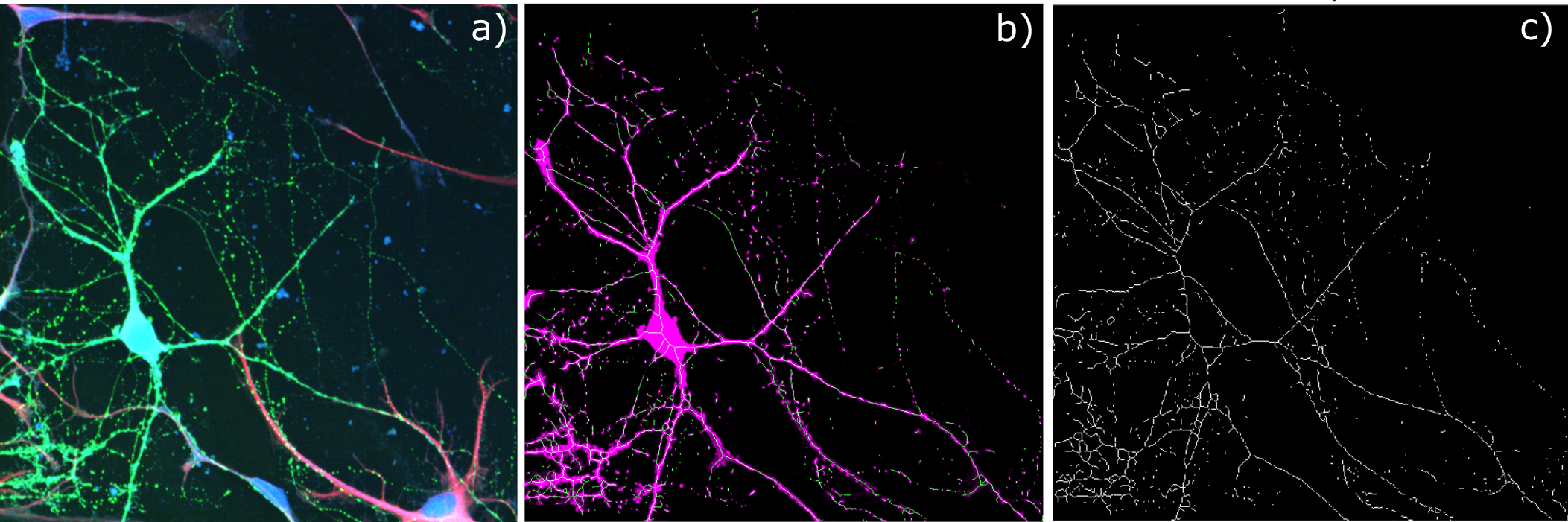}
            \caption{Example Images of the skeletonization process. a) The neuron confocal image, is processed to remove glia fluorescence and then skeletonized, b) where each neurite has a line drawn over it. The result is c) a list of lines, which can be integrated to quantify neuron lengths}
            \label{fig:Ordering}
        \end{figure*}

\subsection{Model Derivation}
To model neuron electrophysiology we are applying the coupled Poisson-Nernst-Planck equations \cite{Lopreore2008a, Pods2013a, Corry2000}: 

\begin{equation}\label{eqnPoisson}
    \epsilon \vec{\nabla^2} V(r) = -\rho(r) = -e \sum_{i = 1}^{M} z_i c_i(r) 
    \end{equation}

\begin{equation}\label{eqnDriftDiffuse}
    \frac{\partial{c_i}}{\partial{t}} = -\vec{\nabla} \cdot [D_i(\vec{\nabla} c_i + \frac{1}{k_b T} z_i e c_i \vec{\nabla} V(r)]  
    \end{equation}
        
The Poisson equation (\ref{eqnPoisson}) utilises the charged ion concentration to solve for the potential and the Nernst Planck equation (\ref{eqnDriftDiffuse}) utilises the electric potential to model ion concentrations in terms of the electrostatic and chemical forces that act on them. In this model, $c_i(r)$ is the ionic concentration, which is proportional to its charge density $\rho(r)$, and $V(r)$ is the electric potential, $k_b$ is the Boltzmann constant, $T$ the temperature, $e$ the electric charge and $z_i$ the ion valency. The increment $i$ denotes which ionic species is being studied (e.g. sodium or potassium), so the total potential will be the solution to the coupled equation, summed over all the participating ion species (up to the total, $M$). Simplifying the model: 

\begin{equation}\label{eqnDDflux}
        \Vec{f}_{\pm} = -D_i\vec{\nabla} c_{i\pm} \mp \mu_i c_{i\pm} \vec{\nabla} V(r) 
        \end{equation}
        
    \begin{equation}\label{eqnRho}
        \rho(r) = -ez(c_{+}-c_{-})  
        \end{equation}

  \begin{equation}\label{eqnJdens}
        \vec{J} = -ez(\vec{f_{+}}-\vec{f_{-}})  
        \end{equation}
        
Where $\vec{f}$ is the flux of a particular ion species, $\mu_i$ is the ion mobility ($\mu_i = \frac{D_i^2k_bT}{ze}$) and $\vec{J}$ is the current density. Although this model can be expanded to include any number of different ion species, for ease of calculation, we only consider the monovalent species: sodium (Na$^+$), potassium (K$^+$), chlorine (Cl$^-$) and negatively charged proteins produced by the neuron (OA$^-$). Although other ions do exist, their concentrations are considered low enough to be neglected. Additionally, the positive ions and the negative ions are grouped into a single averaged source: $c_+ = \frac{Na^+ + K^+}{2}$ and $c_- = \frac{Cl^- + OA^-}{2}$ for computational simplicity. Considering an AP as a travelling wave with constant velocity $v$, we can transform into a moving reference frame to remove the time dependence, allowing the following replacements: 

\begin{equation}\label{eqnTravelwave1}
        \xi = z-vt
        \end{equation}
        
\begin{equation}\label{eqnTravelwave2}
        \frac{\partial}{\partial z} \xrightarrow{} \frac{\partial}{\partial \xi}
        \end{equation}
        
\begin{equation}\label{eqnTravelwave3}        
        \frac{\partial}{\partial t} \xrightarrow{} -v\frac{\partial}{\partial \xi}
        \end{equation}.

To solve the PNP equations, the boundary conditions must be clearly stated and studied for both the ion concentrations and the electric potential. Far from the neuron radially, the boundary conditions are straightforward as the electric potential must go to zero, and the ion concentrations must reflect this with a stable equilibrium: 

\begin{equation}\label{eqnVrboundary}
        V(r)|_{r \xrightarrow{} \infty} = 0 
        \end{equation}
    
\begin{equation}\label{eqncrboundary}
        c_{\pm}(r)|_{r \xrightarrow{} \infty} = c_{b\pm} 
        \end{equation}
        
where $c_{b\pm}$ is the sum of the bulk ion concentrations for the positive and negative ions respectively, which will sum to a zero charge density (see table \ref{table:Parameters}). Axially, far from the AP along the neuron, we expect the ion concentrations and the electric potential to reach a constant equilibrium corresponding to the resting potential (-68 $mV$), thus the derivative of the potential and flux must be zero in this region:

\begin{equation}\label{eqnVzboundary}
        \frac{\partial V(r)}{\partial \xi}|_{\xi \xrightarrow{} \infty} = 0 
        \end{equation}
    
\begin{equation}\label{eqnczboundary}
        f_{\pm}(r)|_{\xi \xrightarrow{} \infty} = 0 
        \end{equation}.

At the membrane, the boundary condition is less clear, however we can derive a boundary condition using the solutions of the Hodgkin-Huxley (HH)\cite{HH1952, Zandt2011a, Lopreore2008a} equations, which state that the total membrane current can be expressed as a sum of components from Kirchoff's laws: 

\begin{equation}\label{eqnHHKirchoff}
        C_m \frac{\partial V(t)}{\partial t} + I_{int}(t)+I_{ion}(t) = 0   
        \end{equation}
        
where the total current per unit area across the membrane is made up of the capacitive current, $C_m \frac{\partial V(t)}{\partial t}$ i.e. ions drifting on and off the membrane, which functions as a capacitor, an initial stimulus current, $I_{int}$ e.g. a current from a distal dendrite, and the ionic current, $I_{ion}(t)$, which can be made up of the sodium and potassium radial currents passing through the membrane during an AP: 

\begin{equation}\label{eqnHHion}
        I_{ion}(t) = I_{Na} + I_{K}
        \end{equation}
        
\begin{equation}\label{eqnHHINa}
        I_{Na}(t) = g_{Na}m(t)^3h(t)(V(t)-V_{Na}) + g_{NaL}(V(t)-V_{Na})
        \end{equation}
        
\begin{equation}\label{eqnHHIK}
        I_{K}(t) = g_{K}n(t)^4(V(t)-V_{K}) + g_{KL}(V(t)-V_{K})
        \end{equation}
    
where $I_{Na}(t)$ and $I_{K}(t)$ are the time dependent sodium and potassium currents respectively. Note that our HH equation has removed any negative ion flux, whose effect on the HH solution is negligible. The currents can be described as the sum of the maximum ion conductance, $g_{Na}$ and $g_{k}$ as well as the leak conductance, $g_{NaL}$ and $g_{kL}$. The conductance terms are multiplied by the overall potential as well as the Nernst potential of each ion species ($V_{Na}$ and $V_{K}$), which can be found using the Nernst equation \cite{Zandt2011a, Hille2001}:

\begin{equation}\label{Nernst}
    \Delta V = \frac{RT}{ezF}ln\Big(\frac{[X]_e}{[X]_i}\Big) 
\end{equation}

where $R$ is the gas constant, $F$ is Faraday's constant, $[X]_i$ is the internal concentration of the ion being studied and $[X]_e$ is the external concentration.

The maximum conductance's are also multiplied by the gating parameters $m(t)$, $h(t)$ and $n(t)$, which are functions that describe how 'open' the gated ion channels are for a specific type of ion channel (sodium or potassium) at a given time. With all these terms, the membrane boundary condition is constructed from the HH equations in the following way: 

\begin{multline}\label{eqnHHFull}
    \frac{dV}{dt} = \frac{1}{C_m}\Big(-I_{int}(t) -g_{Na}m(t)^3h(t)(V(t)-V_{Na}) \\
        - g_{NaL}(V(t)-V_{Na}) \\
        - g_{K}n(t)^4(V(t)-V_{K})
        - g_{KL}(V(t)-V_{K})\Big)    
        \end{multline}

\begin{equation}\label{eqnm(t)}
        \frac{dm}{dt} = \phi(\alpha_m(t)\big(1-m(t)\big)-\beta_m(t)m(t))
        \end{equation}   

\begin{equation}\label{eqnh(t)}
        \frac{dh}{dt} = \phi(\alpha_h(t)\big(1-h(t)\big)-\beta_h(t)h(t))
        \end{equation} 
        
\begin{equation}\label{eqnn(t)}
        \frac{dn}{dt} = \phi(\alpha_n(t)\big(1-n(t)\big)-\beta_n(t)n(t))
        \end{equation}
        
where the $\alpha$, $\beta$ and time constant, $\phi$ terms are found from experimental fitting by Zandt et al. \cite{Zandt2011a}.

\begin{equation}\label{minf}
        m_{\infty}(V) = \frac{\alpha_m (V)}{\alpha_m (V) + \beta_m (V)}
        \end{equation}
        
\begin{equation}\label{alpham}
        \alpha_{m}(V) = \frac{V + 30}{10\big(1-e^{\frac{-(V+30)}{10}}\big)}
        \end{equation}
        
\begin{equation}\label{betam}
        \beta_{m}(V) = 4.0 e^{\frac{-(V+55)}{18}}
        \end{equation}
        
\begin{equation}\label{alphan}
        \alpha_{n}(V) = \frac{V + 34}{100\big(1-e^{\frac{-(V+34)}{10}}\big)}
        \end{equation}
        
\begin{equation}\label{betan}
        \beta_{n}(V) =  0.125e^{\frac{-(V+44)}{80}}
        \end{equation}
        
\begin{equation}\label{alphah}
        \alpha_{h}(V) = 0.07e^{\frac{-(V+44)}{20}}
        \end{equation}
        
\begin{equation}\label{betah}
        \beta_{h}(V) = \frac{1}{1+e^{\frac{-(V+14)}{10}}}
        \end{equation}.

All the parameters in the equations used were found from various literature sources and are displayed in table \ref{table:Parameters}.

The membrane potential boundary condition can be derived from Gauss' law for a cylinder with a radius R, (at the membrane boundary) and an infinitesimal length $d\xi$:

\begin{equation}\label{eqnGauss1}
        2\pi R \vec{E}_{ir}(R)d\xi + 2\pi d\xi \int^{R}_{0} r \frac{\partial \vec{E}_{i\xi}}{\partial \xi} dr = \frac{Q_i d\xi}{\epsilon_r}
        \end{equation}
        
the first term in equation \ref{eqnGauss1} is the integral for the electric field of the length of the cylinder, and the second term is the solution for the electric field at end-caps of the cylinder. Where $\epsilon_r$ is the absolute permittivity of the medium (water), $\vec{E}_{ir}(R)$ and $\vec{E}_{iz}(R)$ are the radial and axial electric fields at the membrane respectively and $Q_i$ is the internal charge per unit length z inside the neuron. Applying Maxwell's laws to change the electric field into a potential: 

\begin{equation}\label{eqnGauss2}
        \frac{\partial V}{\partial \xi}|_{r=R} = -\frac{1}{R}\int^{R}_{0} r \frac{\partial^2 V}{\partial \xi^2} dr - \frac{Q_i}{2\pi R \epsilon_r}
        \end{equation}
        
the first term in equation \ref{eqnGauss2} implies a integral of an electric potential, which is radially uniform. We know that the charge density increases dramatically near the membrane forming a Debye layer \cite{Hille2001, Olivotto1996a}, therefore we know that a radially constant electric field isn't strictly true, however we don't specifically know the deviation from a constant potential so we will make the following approximation: 

\begin{equation}\label{eqngamma}
        \frac{1}{R}\int^{R}_{0} r \frac{\partial^2 V}{\partial \xi^2} dr \approx \frac{-\gamma R}{2}\frac{\partial^2 V}{\partial \xi^2}
        \end{equation} 
        
where $\gamma$ is a factor, which represents the deviation from a radially uniform electric field. The deviation from a constant radial electric field only occurs in the region where the Debye layer has a significant effect on the concentration, approximately 3~nm from the membrane compared to the full 500~nm radius. Thus we predict that the deviation shouldn't be too large ($\gamma = 1$). The second term in equation \ref{eqnGauss2} requires knowledge about the internal charge in the neuron, $Q_i$. This can be derived in the following way: 

\begin{equation}\label{eqnQi}
        Q_i = q_{i0} - v^{-1} 2\pi R\int^{\infty}_{\xi}\Big( \frac{\partial}{\partial \xi} I_{iz} + I_r \Big)d\xi' 
        \end{equation} 
        
where $q_{i0}$ is the total charge per unit length in the resting condition, which is altered during an AP by the radial and axial currents. Similar to the approximation we made in equation \ref{eqngamma}, we approximate the axial current via Ohm's law to change the expression into an electric field, then use another factor, $\eta$, to represent the deviation from a uniform potential, making a quasi-Ohm's law approximation: 

\begin{equation}\label{eqneta}
        Q_i \approx q_{i0} - \eta \pi R^2 \sigma_{i0} v^{-1} \frac{\partial V}{\partial \xi} - v^{-1} 2\pi R \int^{\infty}_{\xi}I_r d\xi' 
        \end{equation} 
        
where $\sigma_{i0}$ is the total internal conductivity per unit length. Putting equations \ref{eqngamma} and \ref{eqneta} into equation \ref{eqnGauss2} gives the Neumann boundary condition for the potential at the neuron membrane: 

\begin{equation}\label{eqnVboundary}
\begin{split}
        \frac{\partial V}{\partial r}|_{r=R} = -\frac{q_{i0}}{2\pi R \epsilon_r} \\
        + \frac{1}{2\pi R \epsilon_r}\Big(\eta \pi R^2 \sigma_{i0} v^{-1} \frac{\partial V}{\partial \xi} + v^{-1}2\pi R \int^{\infty}_{\xi}I_r d\xi'\Big) \\
        - \frac{\gamma R}{2}\frac{\partial^2 V}{\partial \xi^2}
        \end{split}
        \end{equation}
        
However we can alter equation \ref{eqnVboundary} by recognizing that the derivative of the potential with respect to the axial coordinate can be related to time derivative in the HH equation \ref{eqnHHFull}. Making the following substitutions: 

\begin{equation}\label{eqnsub1}
        \frac{\partial V}{\partial t} = \frac{-1}{C} I_{ion}(t) \xrightarrow{} \frac{\partial V}{\partial \xi} = \frac{v^{-1}}{C} I_{ion}(\xi) 
        \end{equation} 
        
\begin{equation}\label{eqnsub2}
\begin{split}
        \int^{\infty}_{t}\frac{\partial V}{\partial t'}dt' \xrightarrow{} -v\int^{\infty}_{\xi} \frac{\partial V}{\partial \xi'} d\xi' \\
        = -v\big(V(\infty) - V(\xi)\big) = \frac{-1}{C}\int^{\infty}_{\xi} I_{ion}(\xi') d\xi'
        \end{split}
        \end{equation} 
        
Where $V(\infty)$ is the resting potential $V_{rest}$, and $I_{ion}$ is equivalent to the radial current $I_r$ Equations \ref{eqnsub1} and \ref{eqnsub2} allows us to substitute the potential derivative and the current integral in equation \ref{eqnVboundary} to obtain a simpler boundary condition whose potential and current terms can be added in from the HH solutions. We also use the HH current solution to calculate the ion flux, which can be used as a membrane concentration boundary condition: 

\begin{equation}\label{eqnVboundaryFinal}
\begin{split}
        \frac{\partial V}{\partial r}|_{r=R} = -\frac{q_{i0}}{2\pi R \epsilon_r} + \frac{1}{2\pi R \epsilon_r}\Big(\eta \pi R^2 \sigma_{i0} v^{-2} C^{-1} I_r(\xi) \\
        + 2\pi R C\big(V_{rest}-V(\xi)\Big) - \frac{\gamma R v^{-1}}{2 C}\frac{\partial I_{r}(\xi)}{\partial \xi}
        \end{split}
        \end{equation}
        
\begin{equation}\label{eqnCboundaryFinalp}
        \frac{\partial c_+}{\partial t}|_{r=R} = \frac{N_A}{e}2\pi R I_r(\xi)
        \end{equation}

\begin{equation}\label{eqnCboundaryFinalm}
        \frac{\partial c_-}{\partial t}|_{r=R} = 0
        \end{equation}
        
where $N_A$ is Avogadro's constant. Due to the assumption that there is only positive ion flow across the membrane, the negative ion flux is zero.

To obtain the magnetic field boundary conditions we apply Ampere's laws where the axial current is treated in the same way as equation \ref{eqneta}.

\begin{equation}\label{eqnAmpere1}
        \vec{\nabla} \times \vec{B} = \vec{J}_{ext}(\xi)     
        \end{equation}
        
Where the external current density is given from equation \ref{eqnJdens}, and the boundary conditions are as follows:

\begin{equation}\label{eqnBfieldr}
        \vec{B}(r \xrightarrow{} \infty) = 0     
        \end{equation}
        
\begin{equation}\label{eqnBfieldz}
        \vec{B}(\xi \xrightarrow{} \infty) = 0     
        \end{equation} 
        
\begin{equation}\label{eqnBfieldmemIz}
        \vec{B}|_{r = R} = -\frac{\mu_0}{2 \pi R} I_z     
        \end{equation}
        
\begin{equation}\label{eqnBfieldmemV}
        \vec{B}|_{r = R} = -\frac{\mu_0}{2} \eta R \sigma_{io} \frac{\partial V}{\partial \xi}     
        \end{equation}

\begin{equation}\label{eqnBfieldmem}
        \vec{B}|_{r = R} = -\frac{\mu_0}{2}\eta R \sigma_{i0} C^{-1} v^{-1} I_r(\xi)     
        \end{equation}
        
\begin{figure*}[!ht]
    \centering
    \includegraphics[width = 0.6\textwidth]{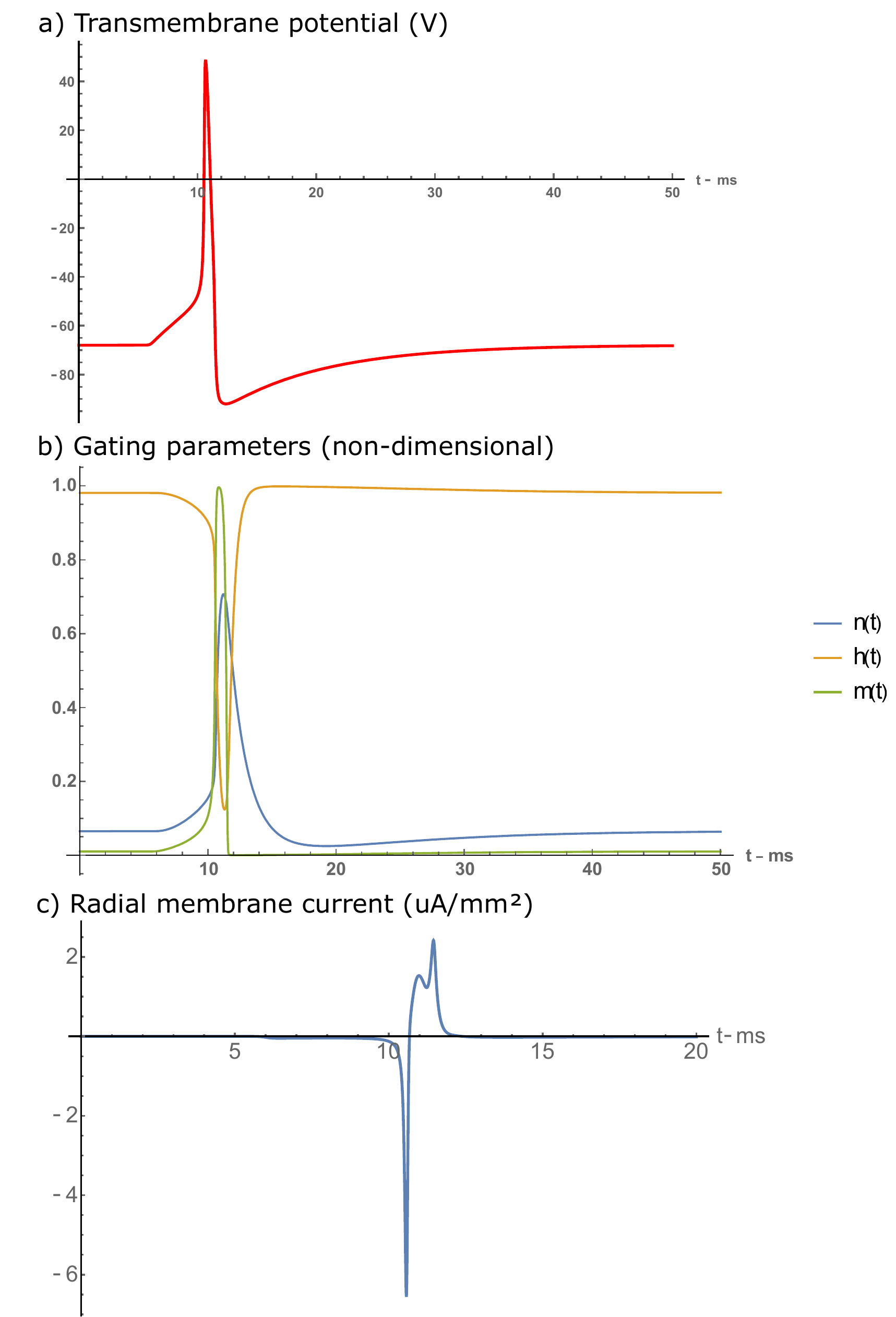}
    \caption{Plots of the a) Hodgkin-Huxley solutions for the potential, b) gating parameters and c) the radial current.}
    \label{fig:HHPlots}
\end{figure*}

\begin{figure*}[!ht]
    \centering
    \includegraphics[width = 0.6\textwidth]{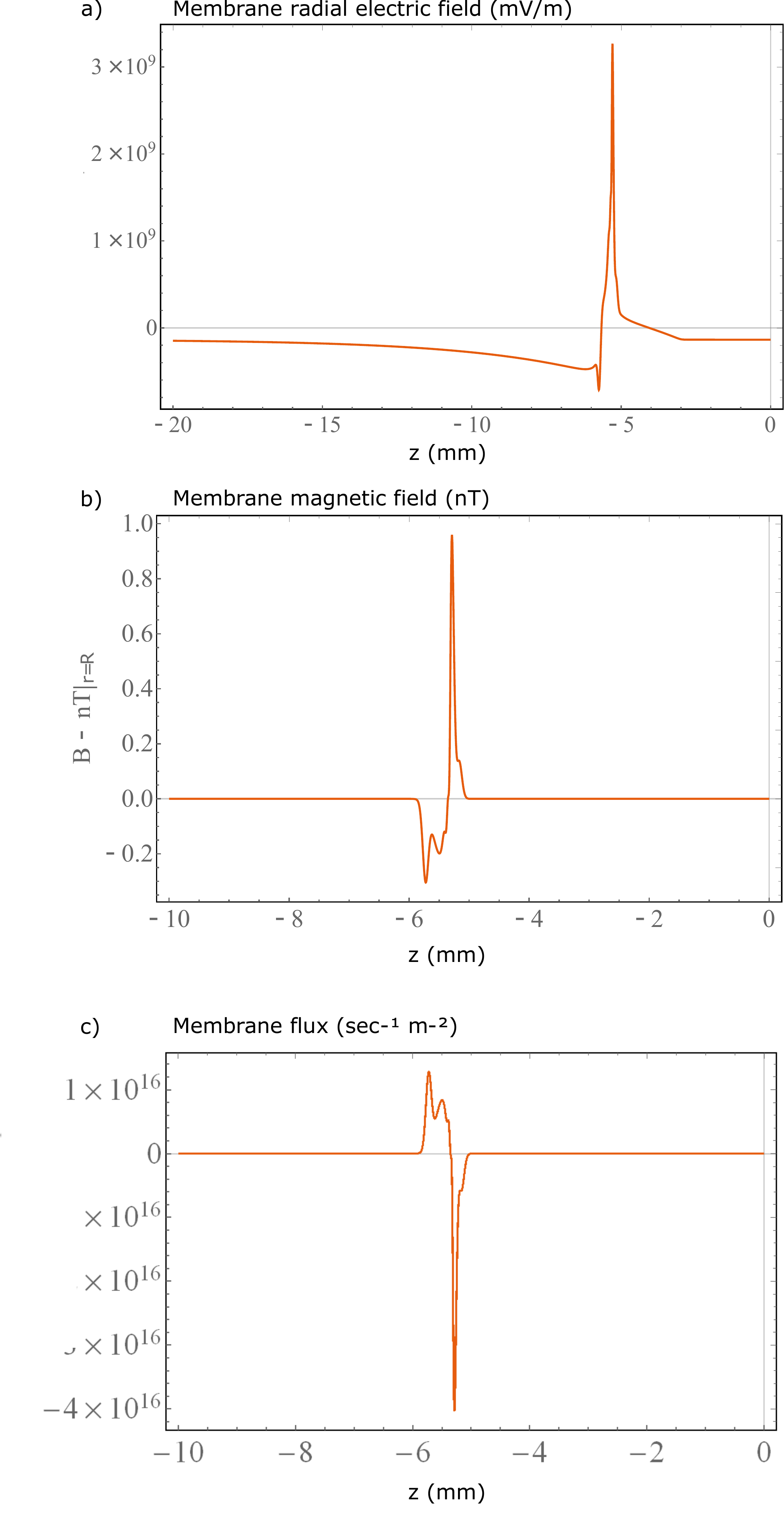}
    \caption{Plots for the a) membrane electric field, b) membrane magnetic field. The plots were derived from equations \ref{eqnVboundaryFinal} and \ref{eqnBfieldmem} respectively. c) Plot of the membrane flux derived from the HH equations. All three of these plots were used as the orange sketch lines in the 2D surface plots of the main paper (figure \ref{Result:2D}).}
    \label{fig:memEBPlots}
\end{figure*}

With the boundary conditions and the equations in place, a 2D axisymmetric cylinder can be constructed whose dimensions are in table \ref{table:Parameters}. The PNP equations are then solved with the stated geometry and boundary conditions using COMSOL Multiphysics. The membrane boundary conditions derived above are featured in figures \ref{fig:HHPlots} and \ref{fig:memEBPlots}. 

\begin{table*}[!ht]
\begin{tabular}{llll}
\hline
\multicolumn{1}{|l|}{Parameter} & \multicolumn{1}{l|}{Description}                      & \multicolumn{1}{l|}{Value}                                         & \multicolumn{1}{l|}{Source}                                        \\ \hline
\multicolumn{1}{|l|}{$K_i$}     & \multicolumn{1}{l|}{Internal Potassium Concentration} & \multicolumn{1}{l|}{155~mmol/L}                                     & \multicolumn{1}{l|}{Lopreore \cite{Lopreore2008a}}   \\ \hline
\multicolumn{1}{|l|}{$K_e$}     & \multicolumn{1}{l|}{External Potassium Concentration} & \multicolumn{1}{l|}{4~mmol/L}                                       & \multicolumn{1}{l|}{Lopreore \cite{Lopreore2008a}}   \\ \hline
\multicolumn{1}{|l|}{$Na_i$}    & \multicolumn{1}{l|}{Internal Sodium Concentration}    & \multicolumn{1}{l|}{12~mmol/L}                                      & \multicolumn{1}{l|}{Lopreore \cite{Lopreore2008a}} \\ \hline
\multicolumn{1}{|l|}{$Na_e$}    & \multicolumn{1}{l|}{External Sodium Concentration}    & \multicolumn{1}{l|}{145~mmol/L}                                     & \multicolumn{1}{l|}{Lopreore \cite{Lopreore2008a}} \\ \hline
\multicolumn{1}{|l|}{$Cl_i$}    & \multicolumn{1}{l|}{Internal Chlorine Concentration}  & \multicolumn{1}{l|}{4.2~mmol/L}                                     & \multicolumn{1}{l|}{Lopreore \cite{Lopreore2008a}} \\ \hline
\multicolumn{1}{|l|}{$Cl_e$}    & \multicolumn{1}{l|}{External Chlorine Concentration}  & \multicolumn{1}{l|}{123~mmol/L}                                     & \multicolumn{1}{l|}{Lopreore \cite{Lopreore2008a}} \\ \hline
\multicolumn{1}{|l|}{$OA_i$}    & \multicolumn{1}{l|}{Internal Protein Concentration}   & \multicolumn{1}{l|}{162.802~mmol/L}                                 & \multicolumn{1}{l|}{Lopreore \cite{Lopreore2008a}} \\ \hline
\multicolumn{1}{|l|}{$OA_e$}    & \multicolumn{1}{l|}{External Protein Concentration}   & \multicolumn{1}{l|}{26~mmol/L}                                      & \multicolumn{1}{l|}{Lopreore \cite{Lopreore2008a}} \\ \hline
\multicolumn{1}{|l|}{$g_{Na}$}  & \multicolumn{1}{l|}{Total Sodium conductance}         & \multicolumn{1}{l|}{100~mS/cm$^2$}                                  & \multicolumn{1}{l|}{Zandt \cite{Zandt2011a}}         \\ \hline
\multicolumn{1}{|l|}{$g_{NaL}$} & \multicolumn{1}{l|}{Sodium leak conductance}          & \multicolumn{1}{l|}{0.0175~mS/cm$^2$}                               & \multicolumn{1}{l|}{Zandt \cite{Zandt2011a}}         \\ \hline
\multicolumn{1}{|l|}{$g_{K}$}   & \multicolumn{1}{l|}{Total Potassium conductance}      & \multicolumn{1}{l|}{40~mS/cm$^2$}                                   & \multicolumn{1}{l|}{Zandt \cite{Zandt2011a}}         \\ \hline
\multicolumn{1}{|l|}{$g_{KL}$}  & \multicolumn{1}{l|}{Potassium leak conductance}       & \multicolumn{1}{l|}{0.05~mS/cm$^2$}                                 & \multicolumn{1}{l|}{Zandt \cite{Zandt2011a}}         \\ \hline
\multicolumn{1}{|l|}{$\phi$}       & \multicolumn{1}{l|}{HH time constant}     & \multicolumn{1}{l|}{3~ms$^{-1}$}                         & \multicolumn{1}{l|}{Zandt \cite{Zandt2011a}}         \\ \hline
\multicolumn{1}{|l|}{$C$}         & \multicolumn{1}{l|}{Membane Capacitance}              & \multicolumn{1}{l|}{1~$\mu$F/cm$^2$} & \multicolumn{1}{l|}{Zandt \cite{Zandt2011a}}         \\ \hline
\multicolumn{1}{|l|}{$T$}         & \multicolumn{1}{l|}{Temperature}                      & \multicolumn{1}{l|}{310$^{o}$K}                                   & \multicolumn{1}{l|}{-}                                             \\ \hline
\multicolumn{1}{|l|}{$_{DK}$}        & \multicolumn{1}{l|}{Potassium Diffusion coefficient}   & \multicolumn{1}{l|}{$1.957\times10^{-9}$~m$^2$/sec}                        & \multicolumn{1}{l|}{Samson \cite{Samson2003a}}       \\ \hline
\multicolumn{1}{|l|}{$D_{Na}$}       & \multicolumn{1}{l|}{Sodium Diffusion coefficient}      & \multicolumn{1}{l|}{$1.334\times10^{-9}$~m$^2$/sec}                        & \multicolumn{1}{l|}{Samson \cite{Samson2003a}}       \\ \hline
\multicolumn{1}{|l|}{$D_{Cl}$}       & \multicolumn{1}{l|}{Chlorine Diffusion coefficient}    & \multicolumn{1}{l|}{$2.032\times10^{-9}$~m$^2$/sec}                        & \multicolumn{1}{l|}{Samson \cite{Samson2003a}}       \\ \hline
\multicolumn{1}{|l|}{$D_{OA}$}       & \multicolumn{1}{l|}{Protein Diffusion coefficient}     & \multicolumn{1}{l|}{$2.00\times10^{-9}$~m$^2$/sec}                         & \multicolumn{1}{l|}{Samson \cite{Samson2003a}}       \\ \hline
\multicolumn{1}{|l|}{$\epsilon_r$}       & \multicolumn{1}{l|}{Absolute Permitivity of water}     & \multicolumn{1}{l|}{$80\times8.854*10^{-12}$~C/Vm}                         & \multicolumn{1}{l|}{-}
  \\ \hline
\multicolumn{1}{|l|}{$V_{rest}$}       & \multicolumn{1}{l|}{Resting Potential}     & \multicolumn{1}{l|}{-68~mV}                         & \multicolumn{1}{l|}{Zandt \cite{Zandt2011a}}       \\ \hline
\multicolumn{1}{|l|}{$R_n$}       & \multicolumn{1}{l|}{Radius of neuron}     & \multicolumn{1}{l|}{500~nm}                         & \multicolumn{1}{l|}{Liewald\cite{Liewald2014a}}       \\ \hline
\multicolumn{1}{|l|}{$L_n$}       & \multicolumn{1}{l|}{axial length of neuron}     & \multicolumn{1}{l|}{2~mm}                         & \multicolumn{1}{l|}{-}       \\ \hline
\multicolumn{1}{|l|}{$R_n$}       & \multicolumn{1}{l|}{Radius nano-mesh}     & \multicolumn{1}{l|}{10~nm}                         & \multicolumn{1}{l|}{-}       \\ \hline
\multicolumn{1}{|l|}{$R_n$}       & \multicolumn{1}{l|}{Radius of external solution}     & \multicolumn{1}{l|}{1.5~$\mu$m}                         & \multicolumn{1}{l|}{-}       \\ \hline
                                &                                                       &                                                                    &                                                                   
\end{tabular}
\caption{Table of Parameters used in the calculations, all other terms used (e.g. ion mobility's or Nernst potentials) are derived from these values. Values that are un-sourced were chosen by the authors to suit the model}
\label{table:Parameters}
\end{table*}

\subsection{Derivation of the electric field in a diamond nanopillar}
\begin{figure*}[!ht]
    \centering
    \includegraphics[width = 0.4\textwidth]{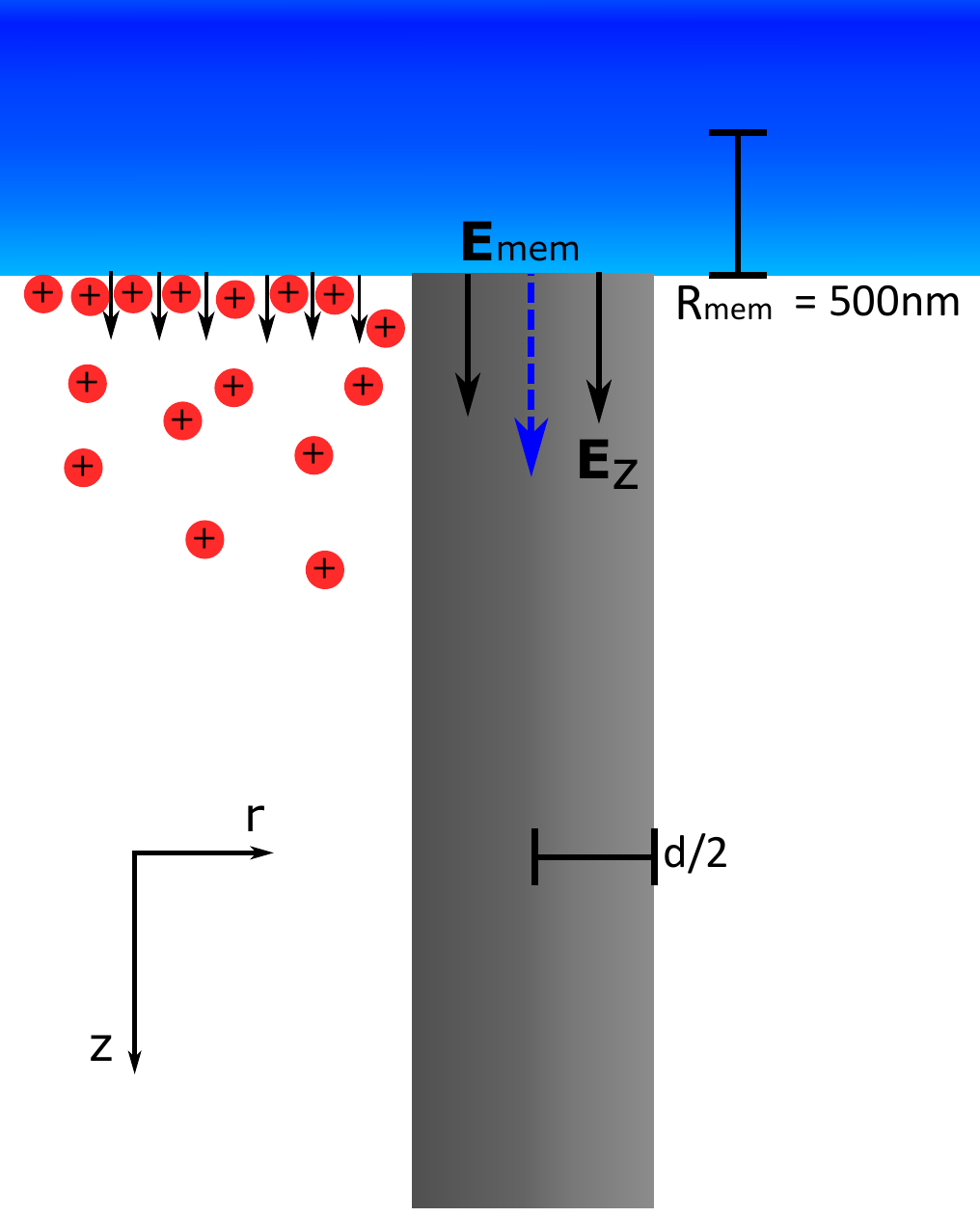}
    \caption{Image of the geometry considered for the electric field inside the diamond when the neurite (blue) runs over the top of diamond pillar (grey). The image features the positive ions forming the Debye layer outside the neuron, which doesn't exist inside the diamond as well as the coordinate system used to obtain the solution. The blue dashed line represents the 1D solution used in figure \ref{Result:1D}b of the main text.}
    \label{Diamondgeometrytip}
\end{figure*}

Consider a geometry where a neuron runs over a single cylindrical diamond pillar where the tip of the pillar is in full contact with the neuron (figure \ref{Diamondgeometrytip}). Assume that the charge inside the neuron is unperturbed by the presence of the pillar and that Debye screening fixes the electric potential on the sidewalls of the pillar to be zero. Note that this ignores the small region close to the neuron (i.e. within the Debye layer) where the potential is non-zero on the sidewalls. We expect these assumptions to be good as long as diameter of the pillar isn't so large that affects the function of the neuron, but sufficiently large compared to the Debye length (1~nm) such that the non-zero potential within the Debye layer has negligible influence on the electric field in the region of the pillar's central axis. This is where it is desirable for the implantation of NV centers. Given these assumptions and adopting the local cylindrical coordinate system of the pillar depicted in figure \ref{Diamondgeometrytip}, Laplace's equation yields the following electric potential within the nanopillar:

\begin{equation}\label{Diamond1}
    V(r, z) = \Big(\frac{d}{k}\Big) E_m J_0\Big(\frac{r k}{(d/2)}\Big)e^{(-k/(d/2))(z-R_{mem})} 
\end{equation}

where $J_0$ is the zeroth Bessel function, $d$ is the diameter of the diamond pillar (200~nm), $R_{mem}$ is the radius of the neurite (500~nm) and $k\approx 2.4$, the first zero solution of the Bessel function. In addition there is $E_m$, the membrane electric field, i.e. the electric field at $z = R_{mem} = 500$~nm, the membrane boundary. This value is calculated from equation \ref{eqnVboundaryFinal} with the same parameters as used in table \ref{table:Parameters} but with a diamond permittivity ($\epsilon  \approx 6$) instead of water. This yields a membrane electric field of $4.54\times10^{10}$~mV/m. It then follows from $\vec{E}=-\vec{\nabla}V$, that the axial electric field inside the pillar is:

\begin{equation}\label{Ez}
    E_z = E_m J_0\Big(\frac{r k}{d/2}\Big) e^{(-k/(d/2))(z-R_{mem})}  
\end{equation}

On the central axis of the pillar where $r=0$, the Bessel function becomes $1$ and the electric field propagation becomes: 

\begin{equation}\label{Ezcentral}
    E_z = E_m e^{(-k/(d/2))(z-R_{mem})}  
\end{equation}

As mentioned in the main text, the axial field decays exponentially from the tip with a decay constant of $k/(d/2)$.

\begin{figure*}[!ht]
    \centering
    \includegraphics[width = 0.6\textwidth]{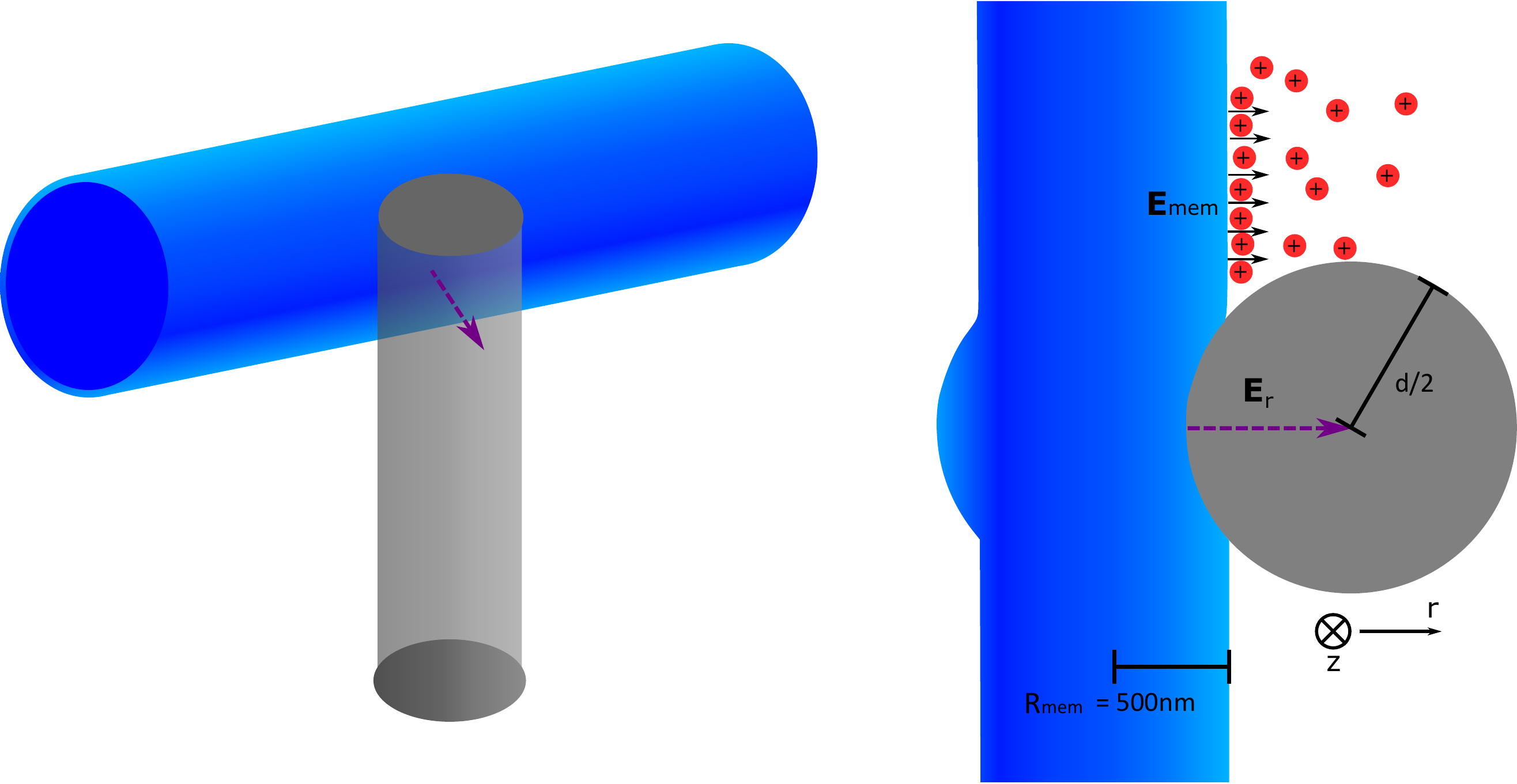}
    \caption{Image of the geometry considered for the electric field inside the diamond when the neurite (blue) runs across the side of diamond (grey) towards the tip of the pillar. The left image shows where upon the diamond pillar the neurite makes contact and the right image shows a top down view of the same system. The image features the positive ions forming the Debye layer outside the neuron, which doesn't exist inside the diamond as well as the coordinate system used to obtain the solution. The purple dashed line represents the 1D solution used in figure \ref{Result:1D}b of the main text.}
    \label{Diamondgeometryside}
\end{figure*}

\begin{figure*}[htp]
    \centering
    \includegraphics[width = 0.65\textwidth]{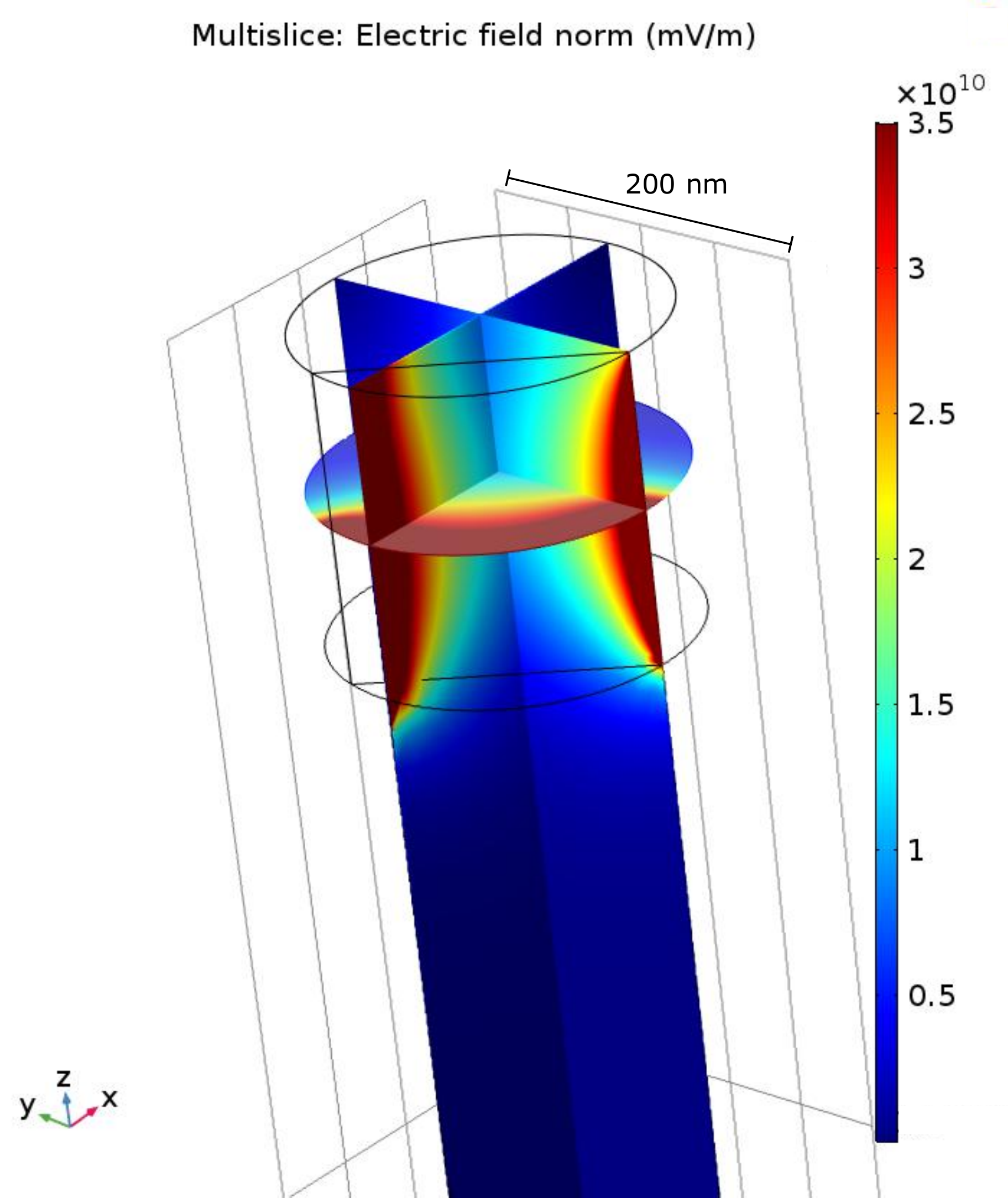}
    \caption{Solution for the electric field with a neurite in contact with the side of the pillar tip. The geometry with the neurite is shown in figure \ref{Diamondgeometryside}. The neurite contact area is marked by the surface of the tip cut out by the black wire-frame rectangle.}
    \label{SideNeuronEField3D}
\end{figure*}

Figure \ref{Diamondgeometryside} depicts the a different modelled geometry, where the neurite runs along the side of the pillar towards the top. In this model, the surface of the pillar, which is in contact with the neurite is the same area as in the case with the neurite running on top of the pillar. This surface portion of the pillar has the same surface electric field from the neuron as the previous case ($4.54\times10^{10}$~mV/m) and the rest of the pillar has its potential fixed at zero. With these boundary conditions the electric field propagation inside the neuron can be solved numerically using COMSOL Multiphysics. The solution in this geometry is shown as a 2D slice density plot in figure \ref{SideNeuronEField3D}.

\clearpage



\end{document}